\documentclass[conference]{IEEEtran}

\usepackage{graphicx}
\usepackage{amsmath}
\usepackage{amssymb}
\usepackage{booktabs}
\usepackage{multirow}
\usepackage{array}
\usepackage[hyphens]{url}
\usepackage{hyperref}
\usepackage{xcolor}
\usepackage{enumitem}
\usepackage{xspace}
\usepackage{balance}
\usepackage{tikz}
\usetikzlibrary{arrows.meta,positioning,fit,calc,backgrounds,shapes.geometric}

\newif\ifsubmit
 \submittrue
\newcolumntype{P}[1]{>{\raggedright\arraybackslash}p{#1}}

\newcommand{\sgx}{\textsc{Intel~SGX}\xspace}
\newcommand{\tdx}{\textsc{Intel~TDX}\xspace}
\newcommand{\sev}{\textsc{AMD~SEV-SNP}\xspace}
\newcommand{\tz}{\textsc{ARM~TrustZone}\xspace}
\newcommand{\cca}{\textsc{ARM~CCA}\xspace}
\newcommand{\ngpu}{\textsc{NVIDIA~H100~CC}\xspace}
\newcommand{\tee}{TEE\xspace}
\newcommand{\tees}{TEEs\xspace}
\newcommand{\cvm}{CVM\xspace}
\newcommand{\cvms}{CVMs\xspace}

\newcommand{\tcb}{TCB\xspace}
\newcommand{\llm}{LLM\xspace}

\newcommand{\rag}{RAG\xspace}
\newcommand{\mcp}{MCP\xspace}
\newcommand{\cc}{CC\xspace}
\newcommand{\mas}{MAS\xspace}
\newcommand{\cmark}{$\checkmark$}
\newcommand{\xmark}{$\times$}
\newcommand{\pmark}{$\sim$}

\newcommand{\marios}[1]{%
  \ifsubmit
  \else
    \noindent{\bfseries \fbox{marios:} {\textcolor{pink}{\itshape #1}}}%
  \fi
}

\newcommand{\javad}[1]{%
  \ifsubmit
  \else
    \noindent{\bfseries \fbox{javad:} {\textcolor{orange}{\itshape #1}}}%
  \fi
}

\begin{document}

\title{When Agents Handle Secrets: A Survey of Confidential Computing for Agentic AI}

\ifsubmit
\author{
\IEEEauthorblockN{Javad Forough}
\IEEEauthorblockA{Department of Computing\\
Imperial College London\\
London, United Kingdom\\
j.forough@imperial.ac.uk}
\and
\IEEEauthorblockN{Marios Kogias}
\IEEEauthorblockA{Department of Computing\\
Imperial College London\\
London, United Kingdom\\
m.kogias@imperial.ac.uk}
\and
\IEEEauthorblockN{Hamed Haddadi}
\IEEEauthorblockA{Department of Computing\\
Imperial College London\\
London, United Kingdom\\
h.haddadi@imperial.ac.uk}
}
\else

\author{
\IEEEauthorblockN{Javad Forough}
\IEEEauthorblockA{Department of Computing\\
Imperial College London\\
London, United Kingdom\\
j.forough@imperial.ac.uk}
\and
\IEEEauthorblockN{Marios Kogias}
\IEEEauthorblockA{Department of Computing\\
Imperial College London\\
London, United Kingdom\\
m.kogias@imperial.ac.uk}
\and
\IEEEauthorblockN{Hamed Haddadi}
\IEEEauthorblockA{Department of Computing\\
Imperial College London\\
London, United Kingdom\\
h.haddadi@imperial.ac.uk}
}
\fi

\maketitle

% -------------------------------------------------------
\begin{abstract}
Agentic AI systems, specifically \llm-driven agents that plan, invoke tools, maintain persistent memory, and delegate tasks to peer agents via protocols such as MCP and A2A, introduce a threat surface that differs materially from standalone model inference. Agents accumulate sensitive context, hold credentials, and operate across pipelines no single party fully controls, enabling prompt injection, context exfiltration, credential theft, and inter-agent message poisoning. Current defenses operate entirely within the software stack and can be silently bypassed by a sufficiently privileged adversary such as a compromised cloud operator. Confidential Computing (CC) offers a hardware-rooted alternative: Trusted Execution Environments (\tees) isolate agent code and data from privileged system software, while remote attestation enables verifiable trust across distributed deployments. This survey synthesizes the design space in four parts: (i)~a unified taxonomy of six \tee platforms (\sgx, \tdx, \sev, \tz, \cca, and \ngpu) covering deployment roles and performance tradeoffs; (ii)~an agent-centric threat model spanning perception, planning, memory, action, and coordination layers mapped to nine security goals; (iii)~a comparative survey of CC-based defenses distinguishing findings that transfer from single-call inference versus what requires new agentic designs; and (iv)~six open challenges including compound attestation for multi-hop agent chains and GPU-\tee performance at \llm scale. While several hardware trust primitives appear mature enough for targeted deployments, no broadly established end-to-end framework yet binds them into a coherent security substrate for production agentic AI.
\end{abstract}

% ============================================================
\section{Introduction}
\label{sec:intro}
% ============================================================

Large language models (LLMs) are increasingly deployed not as isolated,
stateless query-response systems, but as components of \emph{agentic}
systems that plan, invoke external tools, maintain persistent memory,
and act on behalf of users over multi-step time
horizons~\cite{openai_agents_2024, anthropic_mcp_2024}.

\smallskip
\noindent\textbf{The agentic threat surface.}
Moving from stateless inference to autonomous agency changes the
security problem.
An agent accumulates sensitive user context across sessions, holds
service credentials, invokes real-world APIs, and may communicate with
peer agents in pipelines that no single party fully controls.
This expanded operational scope creates a threat surface that
software-layer defenses handle poorly.
A prompt injection attack embedded in a retrieved document can
hijack the agent's planning loop, redirecting its tool calls to
attacker-controlled endpoints~\cite{perez_prompt_injection_2022,
greshake_indirect_2023}.
A compromised cloud operator running the agent's container can
inspect model weights, exfiltrate conversation history, or silently
modify tool outputs in transit.
In multi-agent systems, a single rogue agent can poison the
reasoning of an entire collaborative pipeline~\cite{zhan_injecagent_2024}.

Recent incidents illustrate that these threats are operational rather
than hypothetical.
The 2025 EchoLeak exploit (CVE-2025-32711) demonstrated that a
single malicious email could silently trigger Microsoft Copilot
to exfiltrate sensitive data without any user
interaction~\cite{echoleak_2025}.
Empirical evaluation of unsandboxed agent runtimes has found that
over 75\% of adversarially crafted shell commands execute
successfully against agents lacking runtime
isolation~\cite{agentic_survey_2026}.
Broader recent syntheses at the model and agent levels reach a similar
conclusion: prompt injection, privacy leakage, and unsafe tool use are
emerging as systemic properties of LLM-centered systems rather than
isolated bugs~\cite{yao_llm_security_2024,das_llm_security_2025,
zhang_security_principles_2025}.

\smallskip
\noindent\textbf{The limits of software-only defenses.}
Existing mitigations, including prompt guards, output classifiers,
function-call sandboxes, and agent isolation policies, operate entirely
within the software stack.
Recent examples include graph-based jailbreak filtering and
probabilistic defenses against membership-inference attacks, both of
which harden specific model-level failure modes without changing the
underlying hardware trust boundary~\cite{forough2025guardnet,
forough2025dynanoise}.
They share a common limitation: an adversary with sufficient privilege
below the software boundary can bypass them.
A cloud provider with hypervisor access reads the \llm's weights
and the user's context in plaintext, regardless of application-layer
protections.
A compromised container orchestrator can intercept inter-agent
messages before any application-layer encryption takes effect.
Cryptographic approaches such as Fully Homomorphic Encryption (FHE)
and Secure Multi-Party Computation (MPC) offer rigorous mathematical
guarantees but impose two-to-four orders-of-magnitude overhead,
rendering them impractical for billion-parameter interactive
workloads~\cite{hao_iron_2022, crypten_2021}.
Recent work on private transformer inference sharpens the same tradeoff
in the \llm setting, whether through broader design-space surveys or
newer protocol constructions for secure inference under stronger
privacy constraints~\cite{li_private_transformer_survey_2024,
rathee_mpc_llm_2024,zhang_no_free_lunch_2025,zheng_permllm_2024}.

\smallskip
\noindent\textbf{Confidential computing as a hardware-rooted basis.}
\emph{Confidential computing} (CC) targets this gap:
the absence of hardware-enforced protection for \emph{data in use}.
Trusted Execution Environments (\tees) provide hardware-enforced
isolation of code and data from privileged system software,
including the hypervisor and OS, and enable \emph{remote
attestation}, a mechanism by which a remote verifier can
cryptographically confirm that specific, expected code is executing inside a genuine TEE on a platform with a known security configuration (verified hardware authenticity, firmware version, and patch level),
before sending it any sensitive data.

The hardware landscape has advanced substantially.
\sgx~\cite{costan_sgx_2016} provides process-level enclaves.
\tdx~\cite{intel_tdx_2023} and \sev~\cite{amd_sev_snp_2020}
extend protection to the VM level as Confidential VMs (\cvms).
\tz~\cite{arm_tz_2009} remains widely used in the edge and mobile space.
\cca~\cite{arm_cca_spec_2023} introduces a four-world isolation
model in Armv9-A that reduces the trusted computing base (\tcb)
below the hypervisor level.
\ngpu~\cite{nvidia_h100_cc_2023} extends the trust boundary to
GPU-accelerated inference via Compute Protected Regions.
A 2025 International Data Corporation (IDC) survey of more than 600 global IT leaders across 15 industries found that 75\% of organizations are adopting CC (18\% already in production and 57\% in pilot or testing phases), with reported use cases spanning AI model training, inference, and agent workloads on regulated datasets~\cite{idc_cc_2025}.

\smallskip
\noindent\textbf{The gap.}
Despite this convergence, prior survey work addresses this intersection
only partially rather than through a sustained focus on how CC applies
to the \emph{full agentic stack}.
Existing surveys address agentic AI security from a
threat-modeling perspective~\cite{deng_agent_security_2025,
yu_trustworthy_agents_2025} or CC for standalone LLM
inference~\cite{confidential_llm_survey_2025}, but the
intersection between hardware-rooted trust as a defense substrate for
multi-step, multi-agent autonomous systems remains only partially
explored.
The agentic setting introduces several requirements beyond single-call
inference, including tool-call isolation, persistent memory
confidentiality, message authenticity and provenance across
inter-agent communication, freshness of attestation evidence, and
compositional security across heterogeneous agent chains.

\smallskip
\noindent\textbf{Contributions.}
The paper's main contributions are primarily in synthesis,
comparative analysis, and problem framing:

\begin{enumerate}[leftmargin=*, topsep=2pt, itemsep=1pt]
  \item \textbf{TEE platform taxonomy and comparative review} (\S\ref{sec:bg:tee}
    and \S\ref{sec:platforms}).
    A unified taxonomy and comparative analysis of \sgx, \tdx, \sev,
    \tz, \cca, and \ngpu across six axes directly relevant to agentic deployments:
    isolation granularity, \tcb size, accelerator confidentiality scope (i.e., whether the platform can extend its attested trust boundary into GPU memory and execution, distinguishing NVIDIA H100 CC from CPU-centric TEEs that can use GPUs but cannot protect data in GPU High Bandwidth Memory (HBM)), \marios{How can this be an axis? You have a mix of CPU and GPU TEEs. No TEE prevents GPU use} \javad{Replaced the vague ``GPU support'' axis with ``accelerator confidentiality scope,'' which captures the meaningful distinction: whether a platform can extend its attested trust boundary into GPU memory and execution, as NVIDIA H100 CC does via Compute Protected Regions, versus CPU-centric TEEs that permit GPU use but leave GPU memory unprotected.} edge deployability,
    inter-enclave communication cost, and attestation model, together
    with a balanced discussion of where each platform fits in cloud,
    edge, and accelerator-backed agent deployments.

  \item \textbf{Agent-centric threat model} (\S\ref{sec:threat}).
    A five-layer taxonomy covering perception, planning, memory, action,
    and coordination, mapping each layer to adversarial capabilities,
    concrete threats, and a consistent set of security goals:
    input confidentiality, model confidentiality, execution
    integrity, memory confidentiality, tool-call integrity, message
    authenticity, provenance, freshness, and side-channel resistance.

  \item \textbf{Comparative survey of CC defenses} (\S\ref{sec:tee_agents}
    and \S\ref{sec:multiagent}).
    A structured survey of representative CC-based defenses for LLM and
    agent components, organized via a coverage matrix against our threat
    model and grounded in a broader cited corpus spanning systems
    papers, protocol analyses, surveys, and platform documents.
    Systems analyzed in depth include TEESlice~\cite{teeslice_2024}, CMIF~\cite{cmif_2025}, TEECHAT~\cite{teechat_icml_2025}, Omega~\cite{omega_cloud_agents_2025}, CAEC~\cite{caec_eurosp_2026}, AttestMCP~\cite{attestmcp_2026}, and BlockA2A~\cite{blocka2a_2025} \marios{Add citations. I don't know any of those.} \javad{Added inline citations for all seven named systems; also removed ``among others'' since these are the only systems analyzed in depth---the broader corpus provides context but no additional systems receive this level of treatment.}.

  \item \textbf{Synthesis of open research challenges} (\S\ref{sec:open}).
    A synthesis of six precisely characterized open problems, including
    compound attestation for multi-hop agent chains,
    TEE-backed memory isolation, CC-aware agent communication
    protocols, side-channel leakage in autoregressive inference,
    GPU-TEE overhead at LLM scale, and regulatory alignment under
    the EU AI Act~\cite{eu_ai_act_2024} and DORA~\cite{dora_2022}.
\end{enumerate}

\smallskip
\noindent\textbf{Paper organization.}
\S\ref{sec:method} defines the scope, methodology, and
analytical lens.
\S\ref{sec:threat} develops the agent-centric threat model and its
coverage boundaries.
\S\ref{sec:bg:tee} provides a targeted TEE primer for the hardware
concepts used in later sections.
\S\ref{sec:related} reviews related surveys and positions our work.
\S\ref{sec:tee_agents} surveys CC defenses for core agent components.
\S\ref{sec:platforms} compares platform-specific deployment tradeoffs.
\S\ref{sec:multiagent} addresses multi-agent coordination.
\S\ref{sec:open} identifies open problems.
\S\ref{sec:conclusion} concludes with synthesis takeaways and an outlook on near-term research directions.

\begin{figure*}[t]
    \centering
    \includegraphics[width=\textwidth,trim=0 0 0 0,clip]{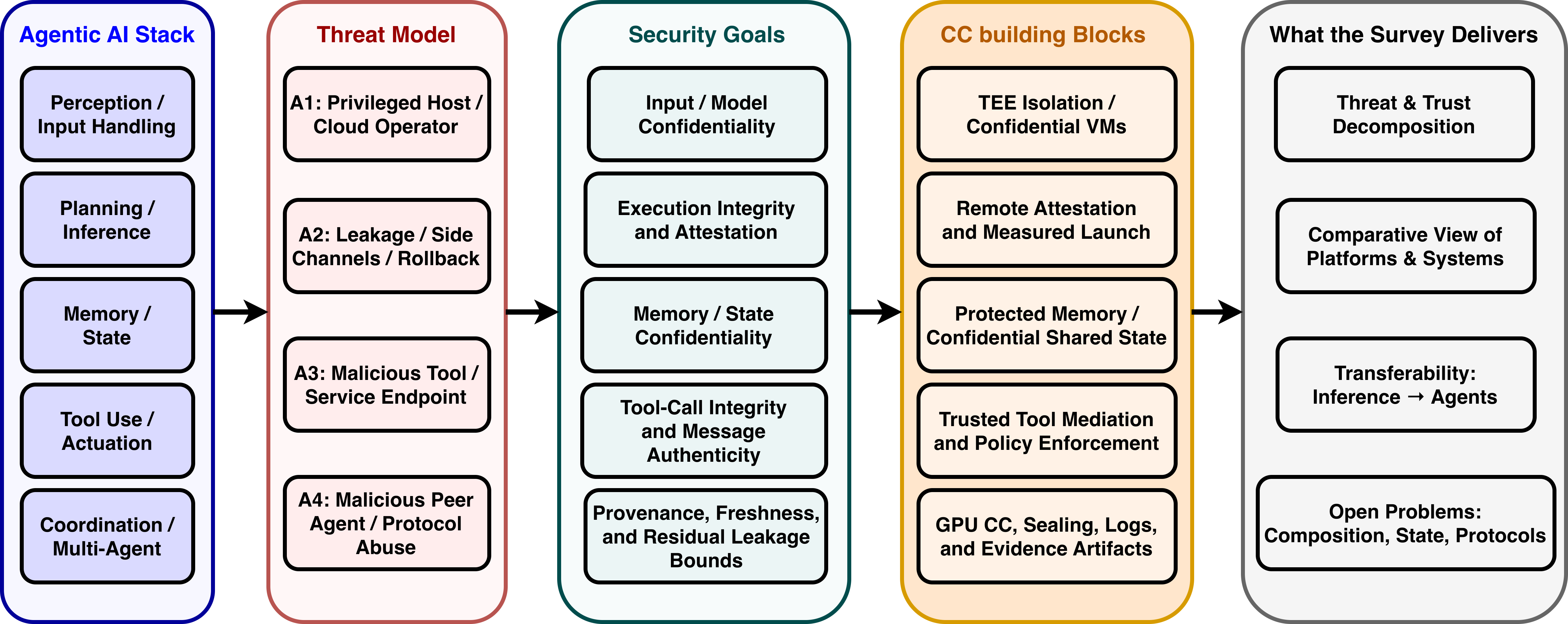}
    \caption{\textbf{Analytical lens of the survey.}
    This figure organizes the paper through five connected abstractions:
    agent components, adversaries, security goals, confidential computing
    mechanisms, and unresolved research gaps. It explains both the
    organization of the survey and the logic behind its comparative
    analysis.}
    \label{fig:survey_lens}
\end{figure*}

% ============================================================
\section{Scope, Methodology, and Analytical Lens}
\label{sec:method}
\label{sec:background}
% ============================================================

\subsection{Scope}

We examine confidential computing as a hardware-rooted defense
substrate for agentic AI against privileged infrastructure adversaries
across cloud, edge, and heterogeneous deployments. The scope is
intentionally narrower than general agent-safety work: we center
hardware isolation and remote attestation, protected memory and tool
execution, and cross-agent trust establishment. Software-only
defenses, alignment methods, and cryptographic computation mechanisms
such as FHE and MPC appear only as comparison points.

The survey uses three analytical dimensions throughout: \emph{functional
agent layer} (perception, planning, memory, action, and coordination),
\emph{adversary privilege} (external attackers, compromised co-tenants,
malicious infrastructure operators, and compromised peer agents), and
\emph{deployment boundary} (process enclaves, confidential VMs,
secure-world or realm-based isolation, and CPU-to-GPU confidential
execution). Figure~\ref{fig:survey_lens} summarizes this lens and
connects it to the threat model, platform mechanisms, and later
synthesis.

We also distinguish four \emph{operational classes}:
\emph{standalone inference systems}, \emph{tool-using single agents},
\emph{workflow agents}, and \emph{multi-agent systems}. This
distinction matters because much of the current evidence base comes
from confidential \llm inference, which informs perception and
planning confidentiality but is only \emph{partially transferable} to
the memory, action, and coordination problems that define the broader
agentic stack. Recent surveys on multi-agent collaboration and
coordination similarly support treating delegation structure and
coordination policy as first-class analytical dimensions rather than as
extensions of single-agent inference~\cite{tran_multiagent_collab_2025,
sun_multiagent_coordination_2025}.

\subsection{Corpus and Classification Protocol}
\label{sec:method:corpus}
\label{sec:method:criteria}
\label{sec:method:mapping}

The paper is a structured survey
whose rigor depends on a transparent corpus and explicit coding rules.
The literature is assembled in three stages: a seed set drawn from
confidential \llm inference, agent-security studies, and primary
platform or protocol documents; backward and forward snowballing; and
targeted keyword searches over terms such as TEE, enclave,
confidential VM, attestation, confidential inference, \llm agent,
\mcp, A2A, tool execution, and multi-agent security. We also track
benchmark work on stateful tool use because it exposes the long-lived
action semantics that distinguish agent security from one-shot
inference evaluation~\cite{lu_toolsandbox_2025}.

The cited literature is divided into a \emph{core analysis corpus} and
a \emph{context corpus}. The core corpus includes work that makes
confidential computing central to the mechanism, addresses at least one
security goal from \S\ref{sec:method:goals}, and maps directly either
to part of the agentic stack or to a clearly transferable precursor
such as confidential \llm inference or protected model state. We
exclude purely software defenses, purely cryptographic approaches
without a hardware trust boundary, classical TEE papers with no direct
bearing on \llm or agent deployments, and papers that lack enough
technical detail for reliable classification. When multiple versions of
the same system exist, we retain the most complete version and consult
earlier ones only for unique measurements or design details. Platform
specifications and vendor documents are used as background sources
unless they themselves expose a concrete security mechanism suitable
for comparison.

Each retained core paper is coded by TEE substrate, operational class,
protected agent layer(s), adversary class, security goals, protected
scope, and transfer status. A system is labeled \emph{directly
agentic} only when it protects actual tool use, workflow execution,
persistent memory handling, or inter-agent coordination; otherwise it
is treated as \emph{partially transferable}. Operational class follows
the most capable behavior explicitly demonstrated, and a security goal
is counted as covered only when the paper provides a mechanism that
directly addresses that goal against the relevant adversary; otherwise
coverage is marked partial. This coding underlies
Table~\ref{tab:threat_coverage} and Table~\ref{tab:design_space} and
keeps the survey conservative about end-to-end capability claims.

\subsection{Security Goal Taxonomy}
\label{sec:method:goals}

To avoid using broad terms such as ``confidentiality,'' ``integrity,''
and ``trust'' too loosely, the remainder of the paper uses nine
security-goal categories.
\textbf{Input confidentiality} protects user prompts, retrieved
documents, and tool inputs before and during model ingestion.
\textbf{Model confidentiality} protects weights, system prompts, and
other proprietary inference artifacts from disclosure.
\textbf{Execution integrity} ensures that the attested code path and
runtime state are the ones intended by the relying party.
\textbf{Memory confidentiality} protects transient and persistent
agent state, including KV cache, conversation history, and vector
stores.
\textbf{Tool-call integrity} ensures that tool invocations and their
parameters are not altered by an infrastructure adversary and are
released only under the intended policy.
\textbf{Message authenticity} ensures that inter-agent or agent-to-tool
messages originate from the claimed sender.
\textbf{Provenance} captures where a message, tool output, or delegated
result came from and which attested components influenced it.
\textbf{Freshness} ensures that attestation evidence, delegation claims,
and returned results are timely rather than replayed or stale.
\textbf{Side-channel resistance} captures resilience to information
leakage through timing, cache, bus, page-fault, or related
microarchitectural signals.

These categories are intentionally narrower than full agent safety.
They describe the concrete security properties that the reviewed CC systems
attempt to provide, and they are used consistently in the threat model,
comparison tables, and system discussions that follow.
They also connect directly to the deployment-assurance question raised
later in \S\ref{sec:open:regulatory}: attestation evidence, provenance,
freshness, and execution-integrity claims are not only runtime
properties, but also the raw material from which auditable assurance
artifacts must be constructed.

\subsection{What Counts as Agentic AI}
\label{sec:bg:agents}

We adopt the following definition, consistent with recent
taxonomies~\cite{deng_agent_security_2025,
yu_trustworthy_agents_2025}:

\begin{quote}
\emph{An agentic AI system is an LLM-driven entity that autonomously
executes multi-step tasks by iterating through a planning loop,
invoking external tools, reading from and writing to persistent
memory stores, and coordinating with peer agents, with limited
or no human intervention between steps.}
\end{quote}

We decompose an agent into five \emph{functional layers} that
serve as the organizing structure of our threat model
(\S\ref{sec:threat}):

\textbf{Perception.}
The agent ingests inputs: user messages, retrieved documents,
tool responses, and inter-agent messages.
These inputs enter the context window and influence the \llm.

\textbf{Planning / Reasoning.}
The \llm core processes accumulated context and produces a plan:
a sequence of tool calls, code generation tasks, or sub-task
delegations to specialist agents (peer agents focused on a narrower task or domain).
Chain-of-thought prompting~\cite{wei_cot_neurips_2022} and
tool-augmented acting paradigms such as ReAct~\cite{yao_react_iclr_2023}
have substantially advanced these planning capabilities, and also enlarge
the attack surface by enabling richer multi-step adversarial sequences.
This is the agent's cognitive core and the primary target of
semantic manipulation attacks.

\textbf{Memory.}
Agents maintain multiple memory tiers: short-term context
(the active context window), long-term vector stores accessed
via Retrieval-Augmented Generation (\rag), episodic memory
(records of past interactions stored and retrieved across sessions), and parametric memory
(model weights and fine-tuned LoRA adapters).

\textbf{Action / Tool Execution.}
Agents invoke external capabilities: web search, code execution,
database queries, API calls, and file system access.
Tool calls are the main mechanism through which agents
affect the real world and constitute some of the more privileged
operations in the agent lifecycle.

\textbf{Coordination.}
In multi-agent systems (\mas), agents communicate and delegate
tasks via structured protocols.
\mcp~\cite{anthropic_mcp_2024}
has emerged as a widely used interface for agent-to-tool
communication, with adoption across OpenAI, Google, Microsoft,
Amazon, and hundreds of third-party providers.
The Agent-to-Agent protocol (A2A)~\cite{google_a2a_2025}
complements \mcp by enabling agent-to-agent delegation across
organizational and platform boundaries.

\begin{figure}[t]
    \centering
    \includegraphics[width=\columnwidth]{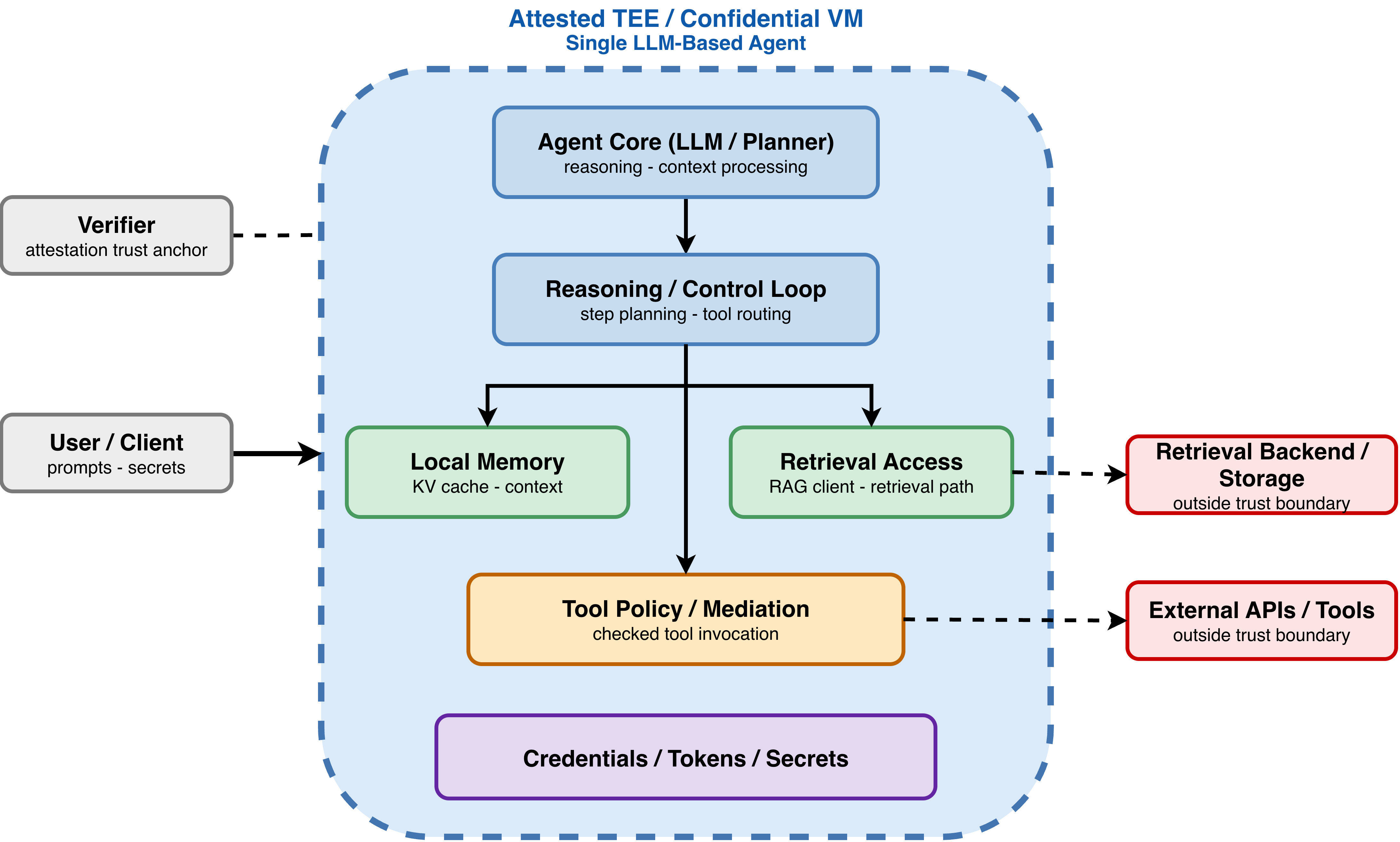}
    \caption{\textbf{Illustrative single-agent trust-boundary model.}
    One attested TEE / confidential VM encloses the LLM-based agent
    core, local memory, credentials, retrieval mediation, and tool-policy
    checks before data or actions leave the runtime. Retrieval backends,
    external storage, APIs / tools, and other I/O paths remain outside
    or only partially covered by the trust boundary.}
    \label{fig:single-agent-tee}
\end{figure}

\begin{figure*}[t]
    \centering
    \includegraphics[width=\textwidth]{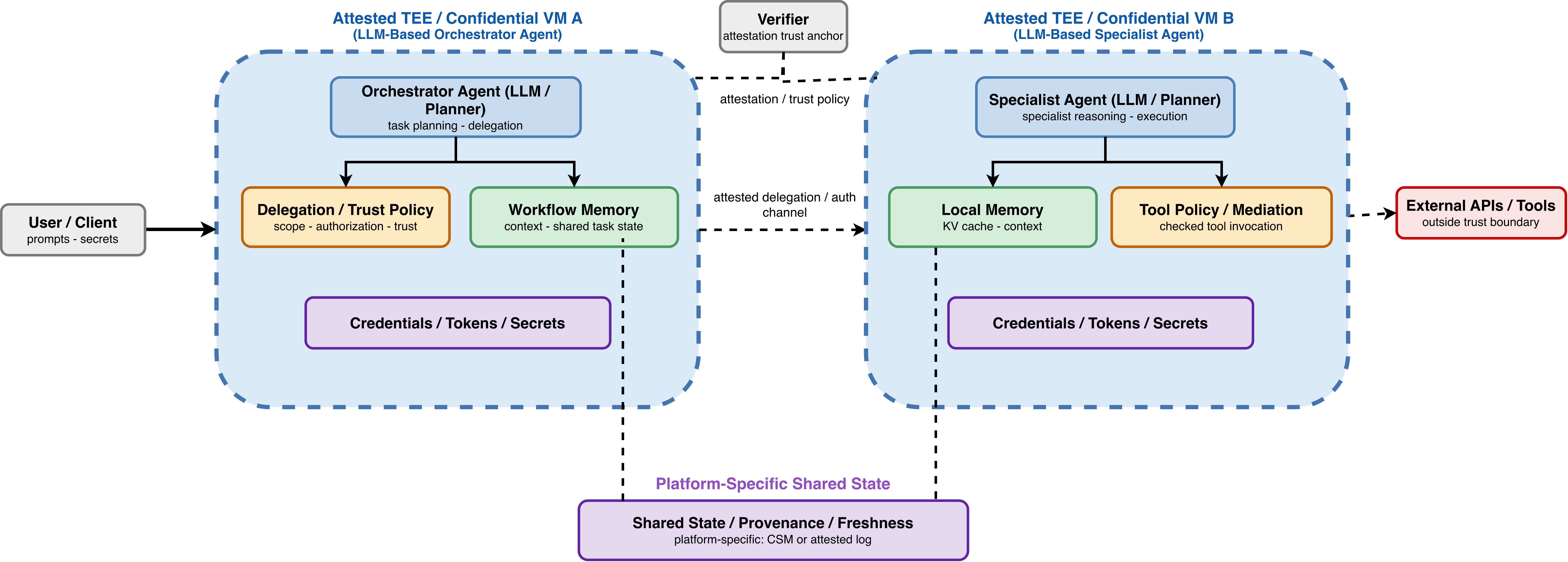}
    \caption{\textbf{Illustrative multi-agent trust-boundary model.}
    Different LLM-based agents execute inside separate attested TEEs /
    confidential VMs, so security depends on remote attestation,
    authenticated delegation channels, and platform-specific shared-state
    mechanisms. Provenance, freshness, and cross-agent policy enforcement
    are conditional properties rather than universally guaranteed.
    External APIs / tools remain outside the trusted boundary.}
    \label{fig:multi-agent-tee}
\end{figure*}

Figures~\ref{fig:single-agent-tee} and \ref{fig:multi-agent-tee}
make these trust-boundary differences concrete.
Figure~\ref{fig:single-agent-tee} presents a single-agent
deployment in which inference, local memory, credentials, and tool
mediation are colocated inside one attested boundary, while external
services and storage remain outside or only partially covered.
Figure~\ref{fig:multi-agent-tee} shows how the security picture changes
once orchestration, delegation, and shared state span multiple
attested components.
The single-agent case is still shaped mainly by whether model
inference, memory, and tool mediation stay within one measured
execution boundary.
The multi-agent case adds a second layer of complexity: each agent may
execute in a different TEE, so confidentiality and integrity depend not
only on local isolation but also on attestation validation,
authenticated communication, freshness, provenance, and the correctness
of cross-agent trust establishment.
Both figures should therefore be read as conceptual trust-boundary
sketches rather than as claims that current systems provide all listed
properties end to end.

\subsection{Targeted TEE Primer}
\label{sec:bg:tee}

Confidential computing protects \emph{data in use} by executing
workloads within hardware-enforced \tees.
For present purposes, only three platform properties matter
up front: the \emph{trust boundary} (process, VM, secure world, realm,
or accelerator region), the \emph{attestation model} by which a remote
party verifies that boundary, and the \emph{deployment envelope} in
which agent components can realistically run.
Accordingly, this primer is deliberately minimal and leaves detailed
platform comparison to \S\ref{sec:platforms}.

At a high level, the platforms discussed later fall into four design
points.
\sgx protects selected application components inside enclaves and is
the narrowest trust boundary considered in this survey.
\tdx and \sev move the boundary to the VM level, making them better
suited to lift-and-shift protection of full agent services, while
retaining a broader \tcb than enclave-based designs.
\tz and \cca represent the Arm lineage: TrustZone provides secure-world
isolation widely used in edge and mobile settings, whereas CCA adds
realm-based isolation for confidential VMs on newer Arm
platforms~\cite{arm_cca_spec_2023}.
\ngpu extends confidentiality into GPU memory and execution, which is
important once model weights, activations, and KV-cache state leave the
CPU boundary~\cite{nvidia_cc_demystified_2025}.

Two primer points matter for the rest of the paper.
First, remote attestation is what turns local isolation into a
distributed trust mechanism: a verifier can decide whether to release
secrets, authorize tool access, or accept delegated results based on
measured code and platform state.
Recent perspective work on confidential computing additionally
emphasizes that attestation evidence is only as useful as the relying
party's ability to interpret both the platform claim and the code
identity being vouched for~\cite{popa_cc_or_crypto_2024,
delignatlavaud_trust_code_2024}.
Second, the boundary chosen for one component does not automatically
cover the rest of an agent pipeline.
Memory services, tool interfaces, CPU--GPU paths, and inter-agent
channels may cross trust boundaries even when inference itself is
protected.
That is why later sections distinguish single-call confidential
inference from end-to-end agent protection, and why platform choice is
best treated as a deployment question rather than as a single abstract
security ranking.

% ============================================================
\section{Agent-Centric Threat Model}
\label{sec:threat}
% ============================================================

\marios{There is one fundamental question that I have and does not get answered there and it's the following. What is different about agents that requires special care? I understand the comparison with stateless LLM inference, but the rest sounds very similar to microservices work or any other cloud application. I think this is key to answer before moving forward.}

\subsection{What Makes Agents a Distinct Security Problem}
\label{sec:threat:distinct}

A natural question is whether agent security reduces to well-understood cloud-service security.
Agents share much of the same infrastructure as microservices: containerized deployment, credential management, inter-service communication, and privileged infrastructure adversaries.
Yet we argue that four structural properties create a security problem that the microservices model handles poorly.

\noindent\textbf{The control plane is semantically manipulable through data.}
In a microservice, application logic is deterministic code whose behavior is fully determined at compile time and is cleanly separated from data it processes.
In an agent, the \llm \emph{is} the control plane: it decides which tools to call, which data to retrieve, and which sub-tasks to delegate, and it makes those decisions based on the content of its input context.
A malicious document embedded in a retrieved corpus can therefore redirect the agent's entire tool-calling and delegation behavior without touching a single authentication boundary.
Prompt injection~\cite{greshake_indirect_2023, perez_prompt_injection_2022, zhan_injecagent_2024, echoleak_2025} exploits exactly this property, and it has no meaningful analog in traditional microservice architectures: there is no call to ``sign your way out'' of a malicious instruction embedded in natural language.

\noindent\textbf{Authorization is implicit, not syntactic.}
Microservices enforce authorization at typed API boundaries: authenticated calls, scoped credentials, and validated schemas ensure that only the intended operations are performed.
An agent's effective authorization, by contrast, is embedded in natural language: in the system prompt, the user instruction, and the accumulated conversation context.
An attacker does not need to bypass authentication; exploiting the fact that the \llm cannot syntactically distinguish a legitimate instruction from an adversarially injected one is sufficient.
The attack surface is the \emph{meaning} of data, not its origin or format.
This shifts the security problem from access-control enforcement to semantic integrity, a property that hardware isolation and attestation cannot directly establish~\cite{perez_prompt_injection_2022, greshake_indirect_2023, attestmcp_2026}.

\noindent\textbf{State is unbounded, accumulating, and long-lived.}
Microservices are largely stateless or operate on well-defined, bounded state machines whose transitions are auditable.
Agents accumulate sensitive context across sessions: KV caches, vector stores, conversation histories, fine-tuned adapters, and retrieved document sets grow indefinitely and without explicit scoping.
The \emph{LeftoverLocals} vulnerability~\cite{leftoverlocal_2024, kv_tee_2024} demonstrated that KV-cache residue in shared GPU memory enables full conversation reconstruction across tenants.
No microservice-style session token or container boundary prevents this because the sensitive data is not a discrete record but a diffuse, high-dimensional representation spread across GPU HBM~\cite{leftoverlocal_2024}.

\noindent\textbf{Multi-agent delegation lacks intent transitivity.}
In a microservice chain, each hop is authorized by explicit, cryptographically scoped credentials: service A calls service B with a token that carries precisely the permissions A was granted, no more.
When an orchestrating agent delegates to a specialist that delegates to a tool server, the user's original intent and scope of authorization exist only as natural language in the context window, a medium that every intermediate agent can silently modify, misinterpret, or be injected into.
No standard mechanism currently proves that the final action executed by the tool server remained within the scope the user originally authorized the orchestrating agent to delegate.
Compound attestation can establish \emph{which code is running} at each hop, but not that the semantic intent of the original delegation was preserved end to end~\cite{omega_cloud_agents_2025, blocka2a_2025, agent_protocol_threats_2026}.

Together, these four properties explain why agentic systems require a distinct threat model rather than an extension of cloud-service security.
The remainder of this section formalizes them into adversary classes, protected assets, and per-layer threat analyses.
Confidential computing addresses the \emph{infrastructure} adversary in this setting, but the semantic and implicit-authorization properties above are why CC alone is insufficient and why each functional layer requires its own analysis.

\javad{Added new subsection directly answering Marios's question about what makes agents distinct from microservices and cloud applications. The four distinguishing properties are: (1) the LLM as a semantically manipulable control plane; (2) implicit, natural-language authorization rather than syntactic enforcement; (3) unbounded accumulating state as a new exposure surface; and (4) multi-agent delegation without cryptographic intent transitivity.}

\subsection{Adversary Model}
\label{sec:threat:adversary}

We consider four adversary classes in increasing strength:

\textbf{A1: External attacker.}
Controls no privileged software.
Can craft malicious inputs (prompt injection via retrieved documents
or tool responses) and observe output side channels.

\textbf{A2: Compromised co-tenant.}
Controls a co-located workload in the same cloud environment.
Can mount cross-VM side-channel attacks (cache timing, memory bus
contention) but does not have hypervisor-level access.

\textbf{A3: Malicious infrastructure operator.}
Controls the hypervisor, cloud management plane, and physical
hardware configuration.
Can read and modify unprotected VM memory, intercept unencrypted
I/O, and manipulate scheduling.
\textit{This is the primary adversary CC is designed to defeat.}

\textbf{A4: Compromised agent.}
One or more agents in a multi-agent pipeline have been
compromised or are inherently malicious.
Can inject false context into shared memory, forge inter-agent
messages, and manipulate orchestration decisions of legitimate peers.

CC-based defenses directly address \textbf{A3}, provide meaningful
protections against \textbf{A2}, and establish attestable trust
boundaries relevant to \textbf{A4}.
They do not prevent the \emph{semantic} effects of prompt
injection (\textbf{A1}): TEE-backed memory isolation can keep
prompts, retrieved context, and runtime state confidential from
a privileged infrastructure adversary, but it does not by itself
stop a prompt-injected agent from leaking that data through its
own outputs or tool calls~\cite{zhan_injecagent_2024, rtbas_2025}.

Throughout the remainder of this section, we describe
coverage using the security-goal taxonomy defined in
\S\ref{sec:method:goals} rather than broad confidentiality or
integrity labels.

\subsection{Protected Assets, Trust Assumptions, and Attack Surface}
\label{sec:threat:assets}

The primary protected assets considered in this survey are the parts of an
agentic system whose disclosure or manipulation would violate the
security goals in \S\ref{sec:method:goals}.
These include: (i) user prompts, retrieved documents, and tool inputs
at the perception boundary; (ii) model weights, system prompts, and
fine-tuned adapters in the planning layer; (iii) runtime state such as
KV cache, conversation history, vector-store contents, and service
credentials in memory; (iv) tool invocations, tool parameters, and
tool outputs at the action boundary; and (v) inter-agent messages,
delegation claims, provenance metadata, and attestation evidence at the
coordination boundary.

Our threat model makes four trust assumptions.
First, the TEE hardware root of trust and its cryptographic
measurement and attestation mechanisms behave according to their
architectural specification.
Second, the measured code inside the attested boundary is the code the
relying party intends to trust.
Third, standard cryptographic primitives used for secure channels,
signatures, and attestation-token validation are not broken.
Fourth, the verifier correctly checks attestation results, freshness,
and policy constraints before releasing secrets or accepting delegated
results.
These assumptions are conventional in the TEE literature, but they are
still assumptions: if they fail, the guarantees discussed in this survey weaken
accordingly.

The main attack surfaces follow directly from the agent lifecycle.
They include the input-ingestion path, model-serving path, persistent
and transient memory state, CPU--GPU transfer paths, storage snapshots,
tool-call interfaces, inter-agent messaging channels, attestation and
replay handling logic, and cross-layer metadata such as timing,
traffic volume, or delegation structure.
Several of these surfaces sit only partially inside the TEE boundary,
which is why the paper repeatedly distinguishes direct coverage from
partial coverage.

Finally, our out-of-scope boundary is intentionally narrow.
We do not claim that CC prevents semantic prompt injection, malicious
but authenticated content, pure denial of service, or misuse that
occurs after data intentionally leaves the attested boundary.
Likewise, invasive physical attacks and microarchitectural leakage
beyond the platform's defended threat model are treated as residual
risks rather than solved properties.
Table~\ref{tab:adversary_assets} summarizes these adversary classes,
the representative assets they place at risk, and the boundary of what
\cc can and cannot guarantee.

\begin{table*}[t]
\caption{Adversary classes, protected assets at risk, and the boundary
of CC guarantees.}
\label{tab:adversary_assets}
\centering
\footnotesize
\renewcommand{\arraystretch}{1.15}
\begin{tabular}{p{1.3cm}p{3.0cm}p{3.0cm}p{3.7cm}p{3.7cm}}
\toprule
\textbf{Class} & \textbf{Representative assets at risk} & \textbf{Adversary capabilities} & \textbf{What CC can help guarantee} & \textbf{What CC does not guarantee} \\
\midrule
\textbf{A1} External attacker & User prompts, retrieved context, tool outputs, planner state influenced by malicious content & Crafts indirect prompt-injection inputs, poisoned retrieval content, malicious tool responses, and observes externally visible behavior & Confidentiality of data that remains inside the attested boundary after ingestion; partial confinement of context against infrastructure-level exfiltration & Semantic correctness, robustness to malicious content, or prevention of unsafe but faithfully executed plans \\
\cmidrule(lr){1-5}

\textbf{A2} Compromised co-tenant & KV cache, GPU residual state, memory-access patterns, timing metadata & Mounts cache, bus, page-fault, contention, or GPU-residual-state attacks without controlling the hypervisor & Partial protection against direct memory disclosure and stronger isolation than conventional VMs or processes & Strong side-channel resistance, metadata hiding, or immunity to all microarchitectural leakage \\
\cmidrule(lr){1-5}

\textbf{A3} Malicious infrastructure operator & Model weights, prompts, credentials, vector stores, tool parameters, inter-agent messages, snapshots & Controls hypervisor, cloud management plane, scheduling, unprotected I/O paths, and storage snapshots & Input, model, and memory confidentiality inside the TEE; execution integrity of measured code; partial tool-call integrity, message authenticity, provenance, and freshness when attestation is composed correctly & Truthfulness of external services, availability, rollback resistance without additional state mechanisms, or protection once data leaves the trust boundary \\
\cmidrule(lr){1-5}

\textbf{A4} Compromised agent & Shared memory, delegated tasks, inter-agent messages, provenance chains, audit records & Injects false context, forges or replays messages, abuses delegation chains, and manipulates collaborator decisions within a multi-agent workflow & Attested identity, authenticated channels, partial provenance, and protected shared state when mutually attested components are used & Trust transitivity by default, semantic honesty of an authenticated peer, or complete end-to-end guarantees across heterogeneous agent chains \\
\bottomrule
\end{tabular}
\end{table*}

\subsection{Per-Layer Threat Analysis}
\label{sec:threat:layers}

\noindent\textbf{Layer 1: Perception.}
Threats: indirect prompt injection via retrieved documents and
tool outputs~\cite{greshake_indirect_2023}; poisoning of external
knowledge sources in the \rag pipeline; eavesdropping on
user inputs in transit or in unprotected memory buffers.
CC coverage: most direct for \emph{input confidentiality} against
infrastructure operators (\textbf{A3}) and partial for
\emph{execution integrity} of the ingestion path.
Semantic injection effects on model behavior (\textbf{A1})
remain outside CC's scope.

\noindent\textbf{Layer 2: Planning / Reasoning.}
Threats: goal hijacking via manipulated context; model weight
extraction by a privileged operator; system prompt
exfiltration; jailbreaking through multi-turn prompt
crafting~\cite{yao_llm_security_2024, das_llm_security_2025}.
CC coverage: most direct for \emph{model confidentiality} and
\emph{execution integrity}.
Remote attestation contributes evidence about code identity and
configuration, and freshness mechanisms determine whether that evidence
is timely~\cite{anati_attestation_hasp_2013, sardar_attestation_2024}.

\noindent\textbf{Layer 3: Memory.}
Threats: unauthorized read of long-term vector stores
by the infrastructure operator (\textbf{A3}); KV-cache leakage
enabling full conversation reconstruction~\cite{leftoverlocal_2024};
cross-user contamination in multi-tenant memory pools;
poisoning of episodic memory by a compromised peer
agent (\textbf{A4}).
CC coverage: most direct for \emph{memory confidentiality} against
\textbf{A3}, but only partial for provenance, freshness, and
multi-tenant isolation of shared memory state.

\noindent\textbf{Layer 4: Action / Tool Execution.}
Threats: leakage of service credentials held by the agent
to a privileged operator (\textbf{A3}); execution of malicious
code generated by the \llm outside an isolated sandbox;
man-in-the-middle modification of tool responses.
CC coverage: most direct for \emph{tool-call integrity},
\emph{execution integrity}, and protection of credentials in memory.
The Omega platform~\cite{omega_cloud_agents_2025} demonstrates
attestation-gated tool access control within a confidential
agent runtime.

\noindent\textbf{Layer 5: Coordination.}
Threats: eavesdropping on inter-agent messages by the
infrastructure operator (\textbf{A3}); agent impersonation
(\textbf{A4}); protocol-level injection vulnerabilities in
\mcp~\cite{attestmcp_2026}; trust transitivity abuse in
multi-hop delegation chains.
CC coverage: most direct for \emph{message authenticity},
\emph{provenance}, and partially for \emph{freshness} when attestation
and replay protection are composed correctly.
CAEC~\cite{caec_eurosp_2026} enables confidential shared memory
between mutually attested CCA realms.
Protocol-level injection (\textbf{A1}) and multi-hop attestation
transitivity (\textbf{A4}) remain open problems
(see \S\ref{sec:open}).

\subsection{Coverage Summary}

Table~\ref{tab:threat_coverage} operationalizes this threat
model as a threat-to-defense coverage matrix.
Rather than only stating which threats exist, it shows how
representative systems map onto the five agent layers and where the
surveyed defenses still leave major gaps.

\begin{table*}[t]
\caption{Threat-to-defense coverage matrix for representative surveyed
systems. \cmark: direct coverage; \pmark: partial or indirect
coverage; \xmark: not addressed.}
\label{tab:threat_coverage}
\centering
\footnotesize
\renewcommand{\arraystretch}{1.2}
\resizebox{\textwidth}{!}{%
\begin{tabular}{p{1.45cm}p{2.05cm}ccccccc}
\toprule
\textbf{Layer} & \textbf{Threat} & \textbf{TEESlice} & \textbf{CMIF}
  & \textbf{TEECHAT} & \textbf{Omega} & \textbf{AttestMCP}
  & \textbf{CAEC} & \textbf{BlockA2A} \\
\midrule
Perception   & Input confidentiality           & \pmark & \cmark & \cmark & \cmark & \xmark & \xmark & \xmark \\
Planning     & Model confidentiality           & \cmark & \pmark & \cmark & \cmark & \xmark & \xmark & \xmark \\
Planning     & Execution integrity             & \pmark & \pmark & \cmark & \cmark & \xmark & \xmark & \xmark \\
Memory       & Memory confidentiality          & \pmark & \pmark & \pmark & \pmark & \xmark & \cmark & \xmark \\
Action       & Tool-call integrity             & \xmark & \xmark & \xmark & \cmark & \cmark & \xmark & \xmark \\
Coordination & Message authenticity            & \xmark & \xmark & \xmark & \cmark & \cmark & \pmark & \pmark \\
Coordination & Provenance                      & \xmark & \xmark & \xmark & \pmark & \cmark & \pmark & \cmark \\
Coordination & Freshness                       & \xmark & \xmark & \xmark & \pmark & \pmark & \pmark & \pmark \\
Cross-layer  & Side-channel resistance         & \xmark & \pmark & \pmark & \pmark & \xmark & \xmark & \xmark \\
\bottomrule
\end{tabular}}
\end{table*}

\subsection{What Confidential Computing Does Not Solve}
\label{sec:threat:limits}

It is equally important to delimit the scope of
confidential computing as to describe its benefits.
CC primarily addresses a subset of the security goals above against a
privileged infrastructure adversary: especially input confidentiality,
model confidentiality, execution integrity, memory confidentiality,
tool-call integrity, message authenticity, provenance, freshness, and
partial side-channel resistance.
It is not a general solution to agent safety, model robustness,
or end-to-end system trustworthiness.

\noindent\textbf{Semantic misalignment and unsafe goals.}
If an agent is poorly specified, misaligned with user intent, or
optimized toward an unsafe objective, a TEE will faithfully execute
that flawed behavior.
Remote attestation can establish \emph{what} code and model are
running, but it cannot establish that their goals are correct,
benign, or aligned with human intent.
The semantic effects of prompt injection, jailbreaks, and deceptive
reasoning therefore remain fundamentally outside CC's protection
boundary.
Recent software-layer work reinforces this point: once agents operate
over stateful tools and multi-turn contexts, prompt-injection resistance
and privacy-leak containment remain essential complementary problems
outside the TEE boundary~\cite{rtbas_2025}.

\noindent\textbf{Trusted-but-malicious or misleading inputs.}
Remote attestation and authenticated channels can authenticate the
measured identity of a tool server, model host, or peer agent, but
attested identity is not the same as truthfulness, safety, or semantic
correctness~\cite{delignatlavaud_trust_code_2024}.
An attested tool may still return harmful instructions, a retrieved
document may still contain adversarial content, and a cooperating
agent may still provide strategically misleading context while
remaining fully within its authenticated software configuration.
In other words, CC can strengthen origin-authentication and
provenance guarantees, but it does not solve input trustworthiness,
content-level verification, or prompt-injection
resistance~\cite{attestmcp_2026, rtbas_2025}.

\noindent\textbf{Availability and denial of service.}
CC is also not an availability mechanism.
A cloud operator can refuse to schedule a confidential VM, throttle
its CPU or GPU resources, delay attestation services, terminate an
enclave, or induce rollback and liveness failures unless additional
protocol safeguards are present.
Indeed, TEE use can introduce new operational fragilities such as
memory-capacity limits, enclave transition overheads, attestation
service dependencies, and device pass-through bottlenecks.

\noindent\textbf{Side channels beyond the architectural model.}
Although TEEs defend against direct memory disclosure by privileged
software, they do not eliminate many microarchitectural and physical
leakage channels.
Timing leakage, cache and bus contention, page-fault or
controlled-channel attacks, speculative-execution effects, GPU
residual-state leakage, and traffic-analysis signals may still expose
information to a sufficiently capable adversary.
This is why side-channel resistance remains a recurring open problem
throughout the paper rather than a solved property of CC platforms.

\noindent\textbf{Model supply-chain compromise.}
Remote attestation establishes that a measured software artifact was
properly instantiated on a genuine TEE / platform with the reported
hardware and security configuration, but it does not by itself
establish the provenance or pre-deployment integrity of that
artifact~\cite{sardar_attestation_2024,
delignatlavaud_trust_code_2024}. A model whose weights were poisoned
during training or fine-tuning, or whose runtime code was backdoored
before measurement, will attest correctly and execute faithfully
inside a \tee while producing adversarially manipulated outputs. CC
therefore provides no protection against supply-chain attacks that
occur before the measured artifact enters the trust boundary.
Defending against this class of threat requires complementary
controls at the model-supply level, including reproducible builds,
model cards, fine-tuning provenance records, and deployment-time
integrity checks that operate outside the TEE's own measurement path.

\noindent\textbf{Security after data leaves the trust boundary.}
Finally, CC only protects data and computation while they remain
inside an attested trust boundary.
Once an agent releases data to an external API, sends a tool command,
stores outputs in an unprotected service, or acts in the physical
world, downstream misuse is governed by protocol design, access
control, sandboxing, and application policy rather than by the TEE
itself.
For agentic systems, this boundary condition is especially important:
hardware-rooted confidentiality is necessary for many deployments, but
it is only one component of a larger safety and security architecture.

% ============================================================
\section{Related Work}
\label{sec:related}
% ============================================================

Prior work falls into four partially
overlapping strands: agent-security surveys, confidential computing
surveys, TEE-focused work on ML and LLM inference, and protocol-level
security analyses for agent communication.
Our goal in this section is to position the paper relative to each
strand and clarify the narrower intersection examined in this survey.

\noindent\textbf{Agentic AI security surveys.}
Several recent works survey security threats in LLM-based
agent systems.
He et al.~\cite{he_llm_agents_2025} survey security and privacy
issues in LLM agents with case studies.
Yu et al.~\cite{yu_trustworthy_agents_2025} examine threats and
countermeasures for trustworthy LLM agents.
Deng et al.~\cite{deng_agent_security_2025} provide a comprehensive
attack-and-defense landscape for agentic AI.
Taken together, these surveys provide the clearest recent treatment of
agent-specific risks such as prompt injection, tool misuse, memory
poisoning, and delegation abuse.
They are therefore important input to our threat model and to the
layered decomposition we use throughout the paper.
Their main emphasis, however, is on attack taxonomies,
software-layer defenses, and trustworthy-agent design rather than on
how hardware-rooted isolation and attestation change the security
picture under privileged-adversary assumptions.

\noindent\textbf{Confidential computing surveys.}
Zobaed and Salehi~\cite{zobaed_cc_survey_2025} survey CC for
machine learning across edge-to-cloud settings.
Feng et al.~\cite{feng_cc_survey_2024} provide a broad review of
CC research.
These surveys provide useful coverage of TEE architectures,
attestation mechanisms, and deployment tradeoffs across the broader CC
landscape.
They also provide much of the hardware and systems context on which our
platform discussion relies.
Their treatment of AI workloads, however, is necessarily broad:
agentic systems do not appear as a first-class organizing category, and
the specific requirements of tool mediation, persistent agent memory,
protocol trust, and multi-agent trust establishment are not central to
their analysis.

\noindent\textbf{TEE-for-ML and LLM inference work.}
Another adjacent line of work studies confidential execution for
machine learning and, more recently, standalone LLM
inference~\cite{confidential_llm_survey_2025}.
Systems such as
CMIF~\cite{cmif_2025}, TEESlice~\cite{teeslice_2024},
TEECHAT~\cite{teechat_icml_2025}, and related systems protect
model weights and user inputs during single-call inference.
Earlier foundational work demonstrated verifiable and private neural
network execution through SGX-offloaded GPU computation~\cite{slalom_iclr_2019},
machine-learning inference as a cloud service over SGX enclaves~\cite{chiron_2018},
selective on-device model protection~\cite{shadownet_sp_2023}, and
cryptographic alternatives spanning oblivious or secret-shared
inference~\cite{gazelle_security_2018, secureml_sp_2017,
delphi_security_2020, cryptonets_icml_2016, cryptflow2_ccs_2020}.
Differential privacy provides a complementary mathematical guarantee for
data release~\cite{dwork_dp_tcc_2006}, while secure aggregation protocols
protect gradient exchange in federated settings~\cite{bonawitz_secagg_ccs_2017,
fltrust_ndss_2021}.
This literature is directly relevant to this survey because it
establishes the state of the art for input confidentiality,
model confidentiality, and parts of execution integrity under TEE
and cryptographic constraints, and it provides early performance evidence for CPU- and
GPU-backed confidential inference.
Our use of it is therefore not as background only, but as precursor
evidence for what does and does not transfer into the broader agentic
setting.
That transfer is strongest for the perception and planning layers and
remains only partial and layer-specific for memory, action, and
coordination.
What remains less developed in prior survey work is a systematic
reinterpretation of these inference-oriented mechanisms against an
explicitly agentic threat model that also includes memory, action, and
coordination layers.

\noindent\textbf{Confidential data systems and TEE infrastructure.}
A body of systems work predating LLM agents has applied SGX and related
TEE hardware to database query processing, distributed analytics, and
data-pipeline execution.
Landmark systems include CryptDB~\cite{cryptdb_sosp_2011} for encrypted
SQL query processing, Opaque~\cite{opaque_nsdi_2017} for oblivious
distributed analytics over SGX, EnclaveDB~\cite{enclavedb_sp_2018} and
VC3~\cite{vc3_sp_2015} for SGX-backed trusted database and analytics
pipelines, StealthDB~\cite{stealthdb_pets_2019} for scalable encrypted
SQL, and Seabed~\cite{seabed_osdi_2016} for analytics over encrypted
datasets.
At the storage layer, SPEICHER~\cite{speicher_fast_2019} and
ShieldStore~\cite{shieldstore_eurosys_2019} demonstrate shielded
key-value stores and LSM-tree engines inside SGX enclaves.
Sandboxed data pipeline execution was demonstrated by
Ryoan~\cite{ryoan_security_2016} for distributed secret-data computation.
Library-OS frameworks such as Haven~\cite{haven_osdi_2014},
Graphene-SGX~\cite{graphene_sgx_atc_2017}, SCONE~\cite{scone_osdi_2016},
and Panoply~\cite{panoply_ndss_2017} reduced the porting burden for
unmodified workloads and remain relevant reference points for lifting
agent runtimes into enclaves.
This earlier body of work is directly relevant to the agentic setting because agent
memory, retrieval paths, vector stores, and tool-facing data stores
require analogous data-in-use protection guarantees; the design patterns
for confidential query processing and data pipeline isolation in SGX are
natural precursors to the persistent-memory and tool-mediation problems
in agentic AI.
The side-channel attack surface of process-level SGX is also well
documented in this era: controlled-channel attacks~\cite{controlled_channel_sp_2015},
the Foreshadow transient-execution attack~\cite{foreshadow_security_2018},
and SgxPectre speculative-execution exploits~\cite{sgxpectre_eurosp_2019}
establish the microarchitectural leakage vocabulary that informs the
open challenge in \S\ref{sec:open:sidechannel}.

\noindent\textbf{Protocol-security work for MCP and A2A.}
Recent work has also begun to analyze the protocol layer directly.
Kong et al.~\cite{kong_mcp_security_2025} focus on security risks in
agent communication protocols, while AttestMCP and related analyses of
\mcp and A2A expose capability-attestation, origin-authentication, and
provenance gaps that become especially important once agents delegate
across organizational boundaries.
This literature sharpens the coordination threat model and motivates
why protocol-level trust establishment matters for agentic systems.
It contributes a complementary perspective to the CC literature by
focusing on protocol semantics, trust propagation, sender identity, and
message provenance.
What it generally does not attempt is a broader synthesis of how those
mechanisms compose with TEE attestation, protected shared state, or
confidential tool execution.

\noindent\textbf{Positioning.}
The agentic-AI setting examined in this survey lies at the intersection of
those four lines of work rather than displacing any one of them.
It uses agent-security surveys to structure the threat model, CC
surveys to ground the hardware and attestation background, TEE-for-ML
and LLM inference work to assess what transfers from confidential inference, and
protocol-security work to analyze coordination trust.
Its incremental contribution is to bring those strands together within
one agent-centric analytical framework: the paper treats agentic AI,
rather than generic AI workloads or protocol design in isolation, as
the organizing problem setting and compares platforms, defenses,
coverage boundaries, and open problems across the full agent stack.

% ============================================================
\section{Survey of CC for Core Agent Components}
\label{sec:tee_agents}
% ============================================================

Table~\ref{tab:design_space} summarizes the design space of the
representative systems analyzed in this section.
The systems differ not only in hardware substrate, but also in which
agent layers they protect, the deployment model they assume, and the
point at which trust is established.
Crucially, the table also marks whether each system is directly
agentic or only partially transferable from standalone inference to
more capable agent settings.
Although the broader cited corpus spans more than one hundred sources, the
analysis in this section emphasizes representative defense systems.
Many sources provide background, measurements, platform context, or
near-variant inference mechanisms, whereas directly agentic and
multi-agent systems remain comparatively few.
The named systems in this section therefore serve as anchors for
recurring design patterns rather than as an exhaustive catalog of the
cited evidence base.
Inference-oriented papers provide the strongest support for
perception-layer and planning-layer protections, with only partial and
layer-specific transfer to memory, action, and coordination.

\begin{table*}[t]
\caption{Design-space summary of representative confidential computing
systems for agentic AI and adjacent LLM settings.}
\label{tab:design_space}
\centering
\footnotesize
\setlength{\tabcolsep}{4pt}
\renewcommand{\arraystretch}{1.15}
\begin{tabular}{P{1.7cm}P{1.85cm}P{2.0cm}P{1.45cm}P{1.8cm}P{1.45cm}P{1.95cm}P{2.1cm}}
\toprule
\textbf{System} & \textbf{TEE substrate} & \textbf{Protected scope} & \textbf{Operational class} & \textbf{Primary agent layers} & \textbf{Agentic status} & \textbf{Main contribution} & \textbf{Main limitation} \\
\midrule
TEESlice~\cite{teeslice_2024} & CPU TEE + GPU offload & Model confidentiality; partial execution integrity & Standalone inference & Perception, Planning & Partially transferable & Shows selective isolation can protect sensitive layers without enclosing the full model & Does not address tool-call integrity, persistent memory confidentiality, or coordination goals \\
\midrule
CMIF~\cite{cmif_2025} & Client-side TEE + remote GPU & Input confidentiality; partial model and memory confidentiality & Standalone inference & Perception, Planning, Memory & Partially transferable & Combines TEE isolation with token-level privacy mechanisms for interactive inference & Protects only part of the stack and leaves tool-call integrity and provenance unaddressed \\
\midrule
TEECHAT~\cite{teechat_icml_2025} & Confidential VM & Input confidentiality, model confidentiality, execution integrity & Standalone inference & Perception, Planning & Partially transferable & Demonstrates near-practical confidential LLM serving at larger scales & Focused on single-agent inference rather than tool-call integrity or message authenticity \\
\midrule
Omega~\cite{omega_cloud_agents_2025} & \shortstack[l]{\textsc{SEV-SNP} \\ \textsc{H100 CC}} & Execution integrity, tool-call integrity, message authenticity & Workflow / tool-using agent & Planning, Action, Coordination & Directly agentic & One of the more complete surveyed steps toward an end-to-end confidential agent runtime & Depends on external protocols and does not fully solve freshness or multi-hop provenance across long delegation chains \\
\midrule
AttestMCP~\cite{attestmcp_2026} & TEE-backed protocol extension & Tool-call integrity, message authenticity, provenance & Tool-using agent middleware & Coordination, Action & Directly agentic & Adds attestation and provenance to MCP-style tool access & Secures protocol establishment, not content semantics or broad memory confidentiality \\
\midrule
CAEC~\cite{caec_eurosp_2026} & \cca realms & Memory confidentiality, message authenticity, partial freshness & Multi-agent substrate & Memory, Coordination & Directly multi-agent & Enables efficient confidential shared state for cooperating agents & Specific to CCA-style shared-memory assumptions and not a general provenance or protocol solution \\
\midrule
BlockA2A~\cite{blocka2a_2025} & TEE + blockchain audit plane & Provenance, auditability, partial freshness & Multi-agent substrate & Coordination & Directly multi-agent & Adds an accountability layer across organizational boundaries & Provides partial provenance support but not full execution integrity or content integrity \\
\bottomrule
\end{tabular}
\end{table*}

\subsection{Practical TEE Limitations for Agentic Systems}
\label{sec:tee_agents:limitations}

Before turning to individual system categories, it is important to make
explicit that TEE-backed protection claims are always conditioned on a
set of practical limitations that are especially salient for agentic AI.
These limitations are not secondary details: they
determine how far claims about inference privacy, memory protection,
and coordinated execution can be sustained in practice.

\noindent\textit{Side channels and controlled leakage.}
Even when a TEE provides strong architectural isolation, inference and
agent runtimes may still leak through timing, cache contention, memory
bus behavior, page-fault patterns, speculative execution effects, or
GPU residual state.
For standalone inference, this weakens claims about model
confidentiality and input confidentiality whenever token-dependent
access patterns are observable.
For agent memory, the same concern extends to KV cache access,
retrieval behavior, and context-length variation.
For multi-agent coordination, communication timing and access
correlations may reveal workflow structure even if message contents are
cryptographically protected.

\noindent\textit{Rollback, replay, and stale state.}
Many TEE deployments provide code identity but weaker guarantees about
state continuity.
An enclave or confidential VM can be reverted to an older snapshot, fed
stale attestation evidence, or induced to consume replayed messages
unless freshness and monotonic-state mechanisms are layered on top.
This limitation is especially serious for agent memory and coordination:
persistent memories, delegated tasks, audit logs, and cross-agent
results all depend on state continuity, not only confidentiality.
Without robust rollback protection, an agent may appear correctly
attested while operating on stale or adversarially replayed state.

\noindent\textit{Accelerator and I/O boundary leakage.}
Modern agents rarely run inside a CPU-only trust boundary.
They depend on GPUs, storage paths, network channels, and tool-facing
I/O stacks.
Whenever those components are only partially covered by the TEE model,
claims about end-to-end execution integrity and memory confidentiality
must be softened.
This is most obvious in confidential LLM inference, where weights and
activations may cross CPU--GPU boundaries, but the same issue appears
for tool outputs, vector-store queries, and shared-memory coordination
primitives.
Accelerator confidentiality therefore improves the picture, but does
not by itself eliminate leakage through driver behavior, DMA paths, or
device scheduling metadata.

\noindent\textit{Metadata leakage and workflow observability.}
Even if message contents remain protected, an infrastructure adversary
may still learn sensitive information from metadata: which tool was
called, when an agent delegated, how many rounds a workflow required,
which peer agents communicated, how large a memory access was, or how
often a retrieval service was queried.
For agentic systems, this can expose business logic, user intent,
organizational structure, or the very existence of a sensitive task.
Accordingly, the confidentiality properties considered in this survey are
content-confidentiality properties unless a system also addresses
metadata minimization or traffic-shaping concerns.

\noindent\textit{Availability limits and operational fragility.}
TEE-based systems remain vulnerable to denial of service, resource
starvation, attestation-service outages, limited secure memory
capacity, expensive enclave or VM transitions, and constrained device
pass-through.
For agents, these are not merely performance annoyances.
They affect whether safety controls can run reliably, whether memory can
remain protected across long sessions, and whether multi-agent
protocols remain live under load.
In practice, some designs that are useful for protecting one
high-value call path may be much less viable for persistent memory or
high-fan-out coordination.

These limitations recur throughout the systems reviewed in this survey.
They are why inference-oriented systems should not automatically be read
as solving agent memory, why attestation is not enough for coordination
without freshness and rollback defenses, and why end-to-end claims in
accelerator-backed deployments depend on more than the CPU-side TEE
alone.

\subsection{Standalone LLM Inference as a Precursor}
\label{sec:tee_agents:inference}

Protecting model inference inside a \tee is a common
starting point for CC-for-AI systems and for the discussion of agentic
security developed in this paper.
At the same time, most systems in this subsection are not agentic in
the full sense used in this survey: they protect a standalone
inference session or chat service rather than an autonomous,
tool-using, multi-step agent.
Their relevance is therefore strongest for the perception and planning
layers, especially \emph{input confidentiality},
\emph{model confidentiality}, and portions of \emph{execution
integrity}, and only indirect for memory, action, and coordination.
Across the reviewed inference literature, the central tension is between the performance requirements of
LLM inference and the memory and compute constraints of
traditional TEE implementations.
This literature is broader than the small set of named systems highlighted in this subsection might suggest, but many papers cluster around a small set of recurring design ideas, so representative examples are more useful than an exhaustive enumeration of near variants.

\noindent\textit{CPU-TEE approaches.}
Early work demonstrated verifiable and private neural network execution
through a complementary design: Slalom~\cite{slalom_iclr_2019}
delegates bulk matrix-multiply operations to an untrusted GPU while a
CPU-side TEE cryptographically verifies the results, achieving near-native
GPU throughput with only modest integrity overhead.
Chiron~\cite{chiron_2018} extended this to a multi-tenant
machine-learning-as-a-service setting, isolating separate clients'
models inside SGX enclaves on shared infrastructure.
Subsequent work demonstrated layer-wise slicing of neural networks
into TEE-resident and GPU-offloaded components.
DarkneTZ~\cite{mo_darknetz_2020} and
T-Slice~\cite{islam_tslice_2023} isolate selected DNN layers
in ARM TrustZone, constraining the subset of layers accessible
to an adversary with memory access.
CMIF~\cite{cmif_2025} deploys the embedding layer inside a
client-side TEE and offloads subsequent computation to GPU
servers, combining hardware isolation with a Report-Noisy-Max-based
token-sanitization mechanism that differentially privately replaces
sensitive tokens with semantically similar alternatives, protecting
sensitive inputs with minimal utility loss.
TEECHAT~\cite{teechat_icml_2025} demonstrates end-to-end private
LLM chat on Azure Confidential VMs, showing negligible overhead
for large models ($>$30B parameters) where inference latency
dominates TEE setup costs.
Recent empirical work on CPU--GPU confidential inference further shows
that the performance question now depends not only on enclave overheads,
but also on how well CPU-side and accelerator-side trust boundaries are
composed in practice~\cite{mohan_cpu_gpu_cc_2024}.
These results transfer most directly to the perception and planning
layers; they do not by themselves establish memory confidentiality,
tool-call integrity, or coordination-layer guarantees for autonomous
agents.

\noindent\textit{GPU-TEE approaches.}
Running billion-parameter models entirely within a CPU-side \tee
is impractical.
The CPU-to-GPU trust extension addresses this by
encrypting data before it traverses the PCIe interconnect between the
CPU TEE and the GPU, where it is placed inside a Compute Protected
Region (CPR) in GPU HBM.
Comprehensive benchmarking across \tdx and \ngpu
platforms~\cite{confidential_llm_perf_2025} reports throughput
penalties of only 4--8\% on GPU \tees, diminishing with
larger batch sizes, a favorable result for multi-tenant agent
deployments.
Again, the evidence is strongest for inference-time input, model, and
memory confidentiality; transfer to tool mediation or coordination
remains indirect.

\noindent\textit{Model partitioning.}
TEESlice~\cite{teeslice_2024} demonstrates that privacy-sensitive
information concentrates in a small, identifiable subset of model
layers.
Isolating only this subset within the \tee, while offloading the
majority to untrusted GPU execution, achieves strong privacy
guarantees at moderate overhead.
This suggests a potentially useful design pattern for agentic
deployments: an agent's system prompt and user context may be processed
inside the trusted boundary while generic reasoning layers execute at
full GPU speed.
The transfer, however, is partial and layer-specific.
It informs perception and planning protection, not memory
confidentiality, tool-call integrity, or coordination goals by itself.

\noindent\textit{Vulnerabilities in partial TEE shielding.}
A 2026 result~\cite{tee_vuln_partial_2026} demonstrates
that a widely adopted optimization, namely precomputed static secret
bases to accelerate cryptographic operations in partial
TEE-shielded inference, introduces a key-reuse vulnerability.
The attack recovers secret permutations and model weights of a
LLaMA-3 8B model in approximately six minutes, scaling to
405B-parameter configurations.
This finding reinforces a recurring theme:
performance-motivated design shortcuts in TEE-LLM integration
can introduce catastrophic security regressions.

\subsection{Protecting Agent Memory}
\label{sec:tee_agents:memory}

Agent memory is among the highest-value targets in the agentic
stack: long-term vector stores accumulate months of proprietary
interactions, fine-tuned adapters encode confidential domain
knowledge, and KV caches contain verbatim conversation
reconstructions.
Here again, the literature is mixed: some mechanisms transfer from
inference-time protection of transient state, but end-to-end protection
of persistent agent memory remains much less mature than protection of
standalone inference.
The dominant security goal in this subsection is therefore
\emph{memory confidentiality}, with provenance and freshness only
partially addressed in current systems.
More broadly, recent surveys of \rag and agentic \rag show that
retrieval is increasingly adaptive, tool-mediated, and long-lived,
which raises the importance of protecting retrieval state and
provenance rather than treating retrieval as a one-shot preprocessing
step~\cite{ding_rag_llms_2024,singh_agentic_rag_2025}.
Confidential database systems provide relevant prior art for protecting
persistent query state: CryptDB~\cite{cryptdb_sosp_2011},
Opaque~\cite{opaque_nsdi_2017}, and StealthDB~\cite{stealthdb_pets_2019}
show how encryption and obliviousness can protect data-at-query-time,
and SPEICHER~\cite{speicher_fast_2019} demonstrates this for shielded
LSM-tree key-value stores, the same storage substrate underlying many
vector databases used by agentic \rag pipelines.

\noindent\textit{KV-cache protection.}
The LeftoverLocals vulnerability~\cite{leftoverlocal_2024}
demonstrated that shared GPU memory allows a co-tenant attacker
to read KV-cache contents from concurrent or previous sessions.
TEE-based KV protection~\cite{kv_tee_2024} demonstrates feasibility
by processing KV pairs inside ARM TrustZone, at the cost of
limiting cache size to the Secure World's constrained memory budget.
CMIF addresses this constraint through a hybrid approach:
noise-based semantic sanitization of sensitive tokens before
they exit the trusted boundary~\cite{cmif_2025}.

\noindent\textit{Multi-tenant vector stores.}
Collaborative Memory~\cite{collaborative_memory_2025} proposes
dynamic access control over shared \rag memory pools, with
per-user permission graphs and memory-confidentiality enforcement
across fully collaborative, asymmetric, and dynamically
evolving access scenarios.
Extending this model to a TEE-backed vector store, where
attestation gates each retrieval operation, is a natural
and currently unrealized combination identified in
\S\ref{sec:open} as an open problem.

\subsection{Tool-Using and Workflow Agents}
\label{sec:tee_agents:tools}

Tool-using and workflow agents are the first directly agentic category
in this section rather than merely inference-oriented.
Here the main protection goals shift from input confidentiality and
model confidentiality for one invocation to execution integrity,
tool-call integrity, and, for delegated workflows, message
authenticity and provenance in a long-lived runtime.
The key goals are \emph{execution integrity}, \emph{tool-call
integrity}, and, once tool servers participate in the protocol,
\emph{message authenticity} and \emph{provenance}.
Tool calls are among the more privileged and higher-risk operations in an
agent's lifecycle.
They cross the boundary from language model output to real-world
effect and are executed with whatever credentials the agent holds.

\noindent\textit{The Omega platform.}
One of the more comprehensive CC-based agent runtimes in the current
literature is
Omega~\cite{omega_cloud_agents_2025}, built on AMD SEV-SNP
and NVIDIA Confidential Compute.
Omega introduces \emph{trustlets}, fine-grained execution units
within a confidential agent runtime, and enforces policy
compliance via a declarative policy language evaluated at the
TEE level before any tool call proceeds.
Agents invoke tools via \mcp and delegate to peers via A2A,
with both channels subject to attestation-gated checks intended to
support tool-call integrity, message authenticity, and provenance.
Evaluation on MCPSecBench shows that, without
CC-backed attestation, MCP-connected tool servers are trivially
impersonable.

\noindent\textit{AttestMCP.}
AttestMCP~\cite{attestmcp_2026} also falls within the tool-using agent category: it extends the \mcp specification with capability attestation and message authentication, providing tool-call integrity and provenance at the protocol-establishment level rather than through a runtime enforcement layer. Because its primary contributions sit at the coordination boundary and are most fully described in the context of multi-agent interaction, its detailed discussion appears in \S\ref{sec:multiagent}. From a tool-call integrity perspective, AttestMCP complements Omega by addressing the protocol layer that Omega's runtime-gated policy model does not directly standardize.

% ============================================================
\section{Platform-Specific Deployment Tradeoffs}
\label{sec:platforms}
% ============================================================

No single TEE platform dominates the agentic AI design space.
\sgx, \tdx, \sev, \tz, \cca, and \ngpu expose different trust
boundaries, programming models, and deployment envelopes, so they are
better treated as complementary than as hierarchically ordered.
This section highlights the deployment niche, main strength, and
principal limitation of each platform in the context of agentic AI.

\subsection{Intel SGX}

\sgx remains the most fine-grained option among widely deployed CPU
\tees.
Its enclave model and narrow \tcb make it attractive for isolating
high-value agent components such as embedding layers, prompt handling,
credential material, and selected memory-management routines.
That same fine granularity, however, makes full-agent execution cumbersome: limited Enclave Page Cache (EPC) capacity, enclave transition overhead, and limited native support for accelerator-rich workloads complicate deployment of large, tool-using agents.
On the accelerator confidentiality scope axis, \sgx cannot extend its attested trust boundary into GPU memory: weights, activations, and KV-cache state that move to the GPU remain outside the \sgx enclave unless a separate GPU \tee (e.g., \ngpu) is introduced alongside it.
Across the reviewed systems, \sgx is best viewed as a platform for protecting
especially sensitive subcomponents rather than entire multi-agent
stacks~\cite{costan_sgx_2016}.
Beyond the agent-facing literature, the SGX ecosystem has produced a
mature body of applied work: verifiable GPU-offloaded neural network
execution~\cite{slalom_iclr_2019}, ML-as-a-service enclaves~\cite{chiron_2018},
on-device model protection~\cite{shadownet_sp_2023},
trusted database query processing~\cite{enclavedb_sp_2018, vc3_sp_2015},
shielded key-value and Log-Structured Merge (LSM)-tree storage~\cite{speicher_fast_2019,
shieldstore_eurosys_2019}, sandboxed data pipelines~\cite{ryoan_security_2016},
and library-OS frameworks for porting unmodified workloads~\cite{haven_osdi_2014,
graphene_sgx_atc_2017, scone_osdi_2016, panoply_ndss_2017}.
Attestation semantics for SGX enclaves are specified in~\cite{anati_attestation_hasp_2013}.
At the same time, the process-enclave model's microarchitectural
attack surface is well documented: controlled-channel
attacks~\cite{controlled_channel_sp_2015}, transient-execution
attacks~\cite{foreshadow_security_2018}, and speculative-execution
attacks~\cite{sgxpectre_eurosp_2019} collectively motivate the
side-channel open challenge in \S\ref{sec:open:sidechannel}.

\subsection{Intel TDX}

\tdx shifts the isolation boundary from process to VM, which aligns
well with contemporary agent deployment practice based on containers,
microservices, and cloud-managed runtimes.
Compared with \sgx, it offers a more practical lift-and-shift path for
full agent services, including orchestration logic, model servers, and
tool adapters.
Its main tradeoff is a broader \tcb and weaker internal compartmentalization:
once code is admitted to the trust domain, the platform itself offers
less structure for separating subfunctions that carry different
confidentiality or execution-integrity requirements.
The emerging TDX Connect path also makes \tdx a relevant CPU anchor for
confidential GPU pipelines, though its end-to-end behavior under
agentic multi-tenant workloads remains under-characterized~\cite{intel_tdx_2023,
intel_tdx_connect_2025}.

\subsection{AMD SEV-SNP}

\sev occupies a similar \cvm-oriented design point to \tdx, and has
already been used as a practical basis for confidential cloud agent
runtimes such as Omega.
Its strengths are deployment maturity, strong VM-level memory
confidentiality, and compatibility with existing virtualization
workflows.
Its most important limitation for agentic AI is that per-VM memory
encryption keys inhibit native confidential shared memory across
co-resident \cvms, which directly affects multi-agent designs that
benefit from shared model state or shared context.
Taken together, the reviewed evidence suggests that \sev is well suited to isolated agent runtimes, but less
naturally suited to tightly coupled multi-agent cooperation on a single
host~\cite{amd_sev_snp_2020, omega_cloud_agents_2025}.
On the accelerator confidentiality scope axis, \sev does not natively extend its trust boundary into GPU memory, but it can serve as the CPU-side attestation root for \ngpu deployments, as demonstrated in Omega, allowing GPU HBM to be included within the combined trust boundary when the two platforms are composed.

\subsection{ARM TrustZone}

\tz remains a widely deployed substrate for mobile and embedded
systems, which gives it continuing importance even though it predates
the recent CC wave.
For agentic AI, its value lies in deployment ubiquity: on-device
assistants, clinical edge devices, and automotive agents are far more
likely to encounter TrustZone-class hardware than server-grade \cvms.
Prior work on selective protection of inference state and KV caches
shows that \tz can secure narrow, high-value parts of an agent pipeline
that primarily carry input confidentiality or memory confidentiality.
That evidence remains mostly inference-oriented rather than a
demonstration of full, tool-using agent runtimes.
Its limitations are equally clear: static world partitioning,
vendor-specific trusted OS stacks, and limited developer-facing
virtualization support make it a constrained platform for general
purpose multi-agent runtimes~\cite{arm_tz_2009}.
On the accelerator confidentiality scope axis, \tz has no native mechanism to extend its secure-world boundary into discrete GPU memory; it is therefore limited to CPU-resident inference and small on-device models where all computation stays within the Secure World.
A systematic survey of prevailing security vulnerabilities in
TrustZone-assisted TEE systems provides important context for how
far these architectural constraints translate to exploitable
weaknesses~\cite{cerdeira_tz_sok_sp_2020}.

\subsection{ARM CCA}

\cca extends Arm's trusted-compute model with realms, a smaller
hypervisor-excluding \tcb, and an attestation path designed for
confidential VMs rather than only secure-world trusted applications.
The reviewed platform and systems literature suggests that these
properties make it relevant to agentic deployments that span
newer Arm server and edge environments.
Research systems such as virtCCA, OpenCCA, SHELTER, CubeVisor, ACAI,
CAGE, and GuaranTEE collectively show that \cca can support
multi-realm coordination, accelerator integration, and on-device
attested inference~\cite{virtcca_2023, opencca_2025,
shelter_usenix_2023, cubevisor_acsac_2024, acai_usenix_2024,
cage_ndss_2024, guarantee_2024, cca_llm_eval_2025,
abdollahi2026agentee}.
Recent platform-level analyses and systems work further clarify this
design point by comparing CCA against TrustZone, formalizing
attestation semantics, and extending realm primitives toward
container-oriented isolation~\cite{garciatobin_cca_2024,
huang_tz_cca_sok_2024,sardar_attestation_2024,zhou_rcontainer_2025}.
On the accelerator confidentiality scope axis, \cca does not natively extend its realm attestation boundary into discrete GPU memory; realm isolation covers CPU-side execution only.
Research prototypes such as ACAI demonstrate that the realm primitive can be extended to on-device accelerators~\cite{acai_usenix_2024}; for server-grade GPU inference, \cca can serve as a CPU-side attestation root alongside a dedicated GPU \tee, analogously to the role \sev plays in \ngpu deployments.
At the same time, it should be treated as one platform among several:
its ecosystem is newer than \sgx, \tdx, and \sev, and many of its
agentic use cases remain demonstrated primarily through recent
research prototypes rather than broad production adoption.

\subsection{NVIDIA H100 Confidential Compute}

\ngpu is qualitatively different from the CPU-centric platforms above:
it extends model confidentiality and memory confidentiality into the
accelerator where LLM inference actually runs.
For large-model agent deployments, this is often difficult to avoid,
because a
CPU-only trust boundary leaves weights, activations, and KV-cache
state exposed once computation moves to the GPU.
Its strengths are performance and relevance to realistic inference
stacks; recent measurements suggest near-native throughput under the
right workload conditions.
Its limitations are equally structural: it depends on a trusted CPU-side
root such as \tdx or \sev for end-to-end system integrity, and it is a
poor fit for edge scenarios where discrete H100-class accelerators are
unavailable~\cite{nvidia_h100_cc_2023, confidential_llm_perf_2025}.
As evidence for agentic deployments, these results are still primarily
about inference-time confidentiality rather than tool-call integrity,
message authenticity, provenance, or freshness.

\subsection{Comparative Takeaway}

\begin{figure}[t]
    \centering
    \includegraphics[width=\columnwidth]{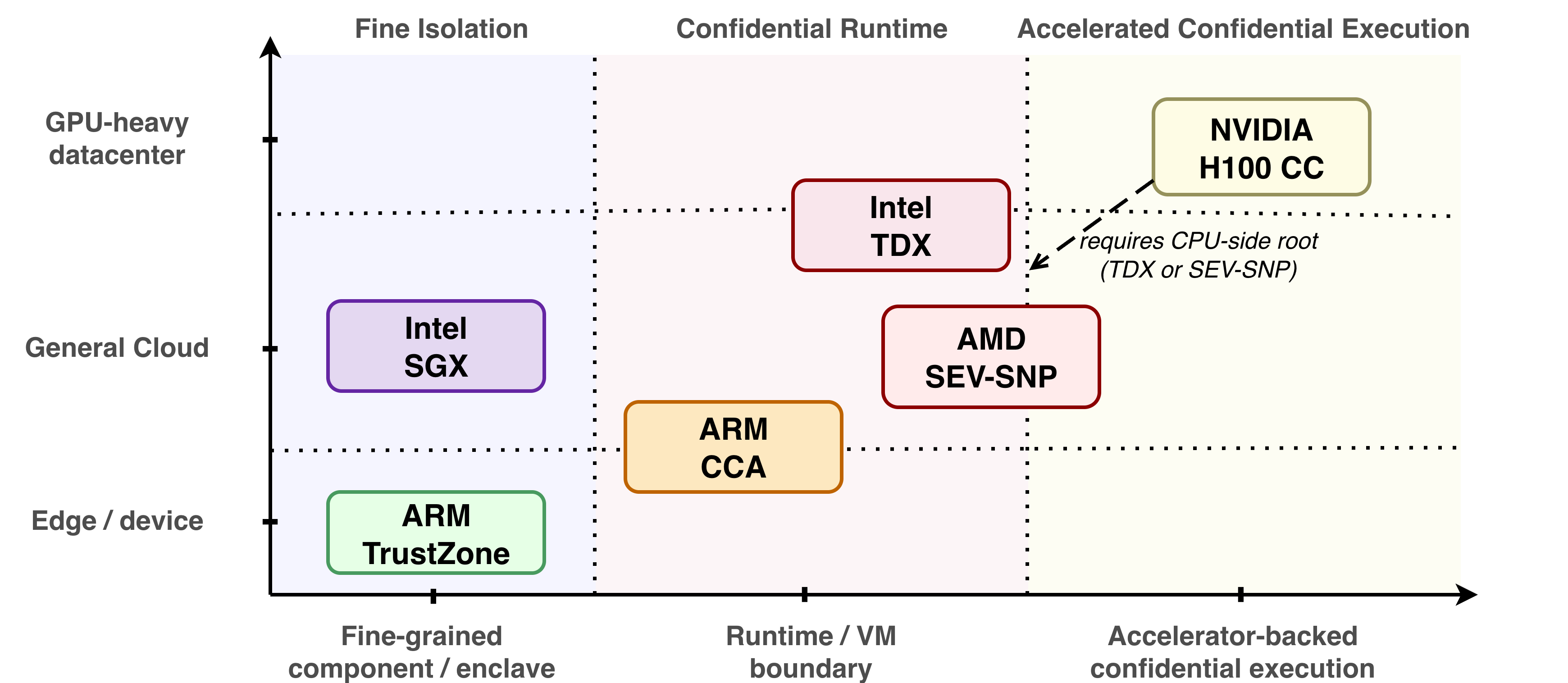}
    \caption{\textbf{Platform landscape for confidential computing in agentic AI.}
    The surveyed platforms occupy different regions of the deployment space rather than forming a simple best-to-worst ranking. Fine-grained TEEs such as Intel SGX are often attractive for small trusted components, confidential-runtime platforms such as Intel TDX and AMD SEV-SNP are better suited to deployable cloud runtimes, and accelerator-backed confidential execution such as NVIDIA H100 CC is relevant for large-model inference in GPU-heavy datacenter settings. ARM TrustZone remains most natural for edge and device deployments, while ARM CCA represents an emerging Arm realm/CVM model spanning edge and cloud use cases. Positions are illustrative and qualitative, emphasizing typical deployment niches rather than a quantitative ranking or maturity ordering. Accelerator-backed platforms such as NVIDIA H100 CC typically rely on a CPU-side confidential computing root (e.g., TDX or SEV-SNP) for end-to-end system trust and do not act as standalone trust anchors.}
    \label{fig:platform_landscape}
\end{figure}

Taken together, the platform literature suggests a
deployment-oriented division
of labor rather than a single winner.
\sgx offers narrow protection for especially sensitive components;
\tdx and \sev support cloud-scale confidential agent services; \tz and
\cca are relevant for edge and heterogeneous Arm deployments; and
\ngpu is a useful enabler for confidential inference at practical LLM
scale.
That broader pattern suggests that agentic AI security
depends on matching the platform to the deployment context and then
composing CPU, VM, edge, and accelerator boundaries so that input
confidentiality, model confidentiality, memory confidentiality,
execution integrity, and coordination-layer goals are covered where
they actually arise. Figure~\ref{fig:platform_landscape} provides a
compact qualitative view of that deployment-oriented positioning.

\subsection{Cross-Cutting Design Lessons}

Across the systems and platforms surveyed in this paper, five recurring design
lessons emerge.

\noindent\textbf{Protect the highest-value layer, not necessarily the whole stack.}
The more practical designs do not attempt to place an entire agentic
system wholesale inside one TEE.
Instead, they align the trust boundary with the layer carrying the
highest-value security goal, such as input confidentiality and model
confidentiality in perception and planning, memory confidentiality for
long-lived state, or tool-call integrity at the action boundary.

\noindent\textbf{Attestation is necessary but insufficient without freshness and provenance.}
Remote attestation is the entry point for execution integrity, but by
itself it does not resolve replay, rollback, or ambiguity about which
attested component influenced a delegated result.
In practice, coordination-layer protections depend on combining
attestation with freshness and provenance rather than treating
measurement alone as a complete trust mechanism.

\noindent\textbf{Inference protection transfers poorly to memory and coordination.}
The strongest current evidence base remains concentrated in
inference-oriented systems, where input confidentiality, model
confidentiality, and portions of execution integrity are most mature.
That evidence transfers only partially to persistent memory
confidentiality, tool-call integrity, message authenticity, and
coordination-layer provenance.

\noindent\textbf{Accelerator trust must be composed with CPU- and VM-level trust.}
GPU confidentiality materially improves model confidentiality and memory
confidentiality during LLM inference, but it does not create an
end-to-end trust boundary by itself.
Practical agent deployments still depend on how CPU-side attestation,
VM isolation, I/O handling, and tool-facing paths compose with the
accelerator boundary.

\noindent\textbf{Shared-state multi-agent designs depend heavily on platform semantics.}
Confidential shared memory is not a generic capability that transfers
uniformly across TEE families.
Whether shared state can support memory confidentiality, message
authenticity, and freshness in a multi-agent system depends strongly on
platform-specific semantics such as keying models, sharing primitives,
and attestation mechanisms.

% ============================================================
\section{Multi-Agent Coordination and Confidential Computing}
\label{sec:multiagent}
% ============================================================

Multi-agent systems should be separated from both standalone
inference and single-agent tool use.
Once multiple agents delegate, attest, and exchange state across trust
boundaries, coordination becomes a problem of message authenticity,
provenance, freshness, and sometimes memory confidentiality in its own
right.
This is also the category where the literature is thinnest and where
transfer from inference-only systems is least reliable.
The main goals in this section are \emph{message authenticity},
\emph{provenance}, \emph{freshness}, and, in some deployments,
\emph{memory confidentiality} for shared state.
The coordination layer is where some of the least studied security
challenges arise and where CC's current coverage is most
incomplete.
Accordingly, the discussion in this section centers on a
small number of representative systems and protocol analyses, because
the evidence base remains limited.

\subsection{Protocol-Level Vulnerabilities in MCP and A2A}

A recent systematic protocol-level security analysis of
\mcp~\cite{attestmcp_2026} identifies three fundamental
specification weaknesses: (i) absence of capability attestation,
allowing any MCP server to claim arbitrary permissions;
(ii) bidirectional sampling without origin authentication,
enabling servers to inject messages into the \llm's context
in the user role without the host distinguishing them from
legitimate user input; and (iii) implicit trust propagation
in multi-server configurations, where the context window
conflates outputs from all connected servers without provenance.
Empirical evaluation across 847 attack scenarios shows that
MCP's architectural choices amplify prompt injection success
rates by 23--41\% relative to equivalent non-MCP integrations~\cite{attestmcp_2026}.

AttestMCP extends the \mcp specification with backward-compatible
capability attestation and message authentication, and combining
this with TEE attestation of the MCP server itself strengthens
tool-call integrity, message authenticity, and provenance at the
tool-provider boundary.
A complementary analysis of \mcp, A2A, Agora, and
ANP~\cite{agent_protocol_threats_2026} proposes context
confidentiality, context integrity, and context availability
as more appropriate security primitives than the traditional
Confidentiality, Integrity, and Availability (CIA) triad for dynamic multi-agent environments.

\subsection{Inter-Agent Trust via Attestation}

Omega~\cite{omega_cloud_agents_2025} establishes one of the more
complete models in the current literature for attested identity, message
authenticity, and provenance across multi-agent exchange: agents, base models,
LoRA adapters, and tools are treated as untrusted by default,
requiring attestation-gated measurement before influencing
execution.
The resulting attested agent identity, a cryptographic
commitment to the full software configuration, is included
in inter-agent messages, enabling peers to verify
collaborators before admitting them to a shared pipeline.

BlockA2A~\cite{blocka2a_2025} complements TEE attestation
with blockchain-anchored audit logs and smart contract-based
capability revocation, creating a two-layer accountability
substrate: TEEs attest \emph{which code is running}; the ledger records \emph{what was done}.

\subsection{Confidential Shared Memory for Multi-Agent Systems}

CAEC~\cite{caec_eurosp_2026}
introduces Confidential Shared Memory (CSM) within CCA's
attested isolation model.
CSM regions are accessible only to mutually attested realms
that have established a shared attestation anchor.
CAEC achieves up to a $209\times$ reduction in CPU cycles
relative to encryption-based sharing through hypervisor-mediated
memory, and sharing an LLM between two CCA realms reduces the
system's total memory footprint by 16.6--28.3\%, an important
scalability result for multi-agent systems running large models
on edge hardware~\cite{caec_eurosp_2026}.

Notably, CAEC-style CSM is architecturally infeasible without
hardware modifications on \sev, whose per-VM encryption keys
preclude confidential cross-\cvm sharing.
Within the current literature, this makes CCA-based designs one of the
clearest current examples of hardware-supported confidential shared
memory for multi-agent deployments that require shared model weights or
shared context under mutual attestation, rather than evidence of a
general platform ranking across all agent settings.

% ============================================================
\section{Open Research Challenges}
\label{sec:open}
% ============================================================

The preceding analysis highlights six open challenges representing the principal
gaps between current CC capabilities and the requirements of
production agentic deployments.

\subsection{Compound Attestation for Multi-Hop Agent Chains}
\label{sec:open:attestation}

Current attestation frameworks are designed primarily to provide
bilateral evidence of execution integrity and freshness:
one verifier attests one enclave.
Production agent pipelines are multi-hop: user $\to$ orchestrator
$\to$ specialist agents $\to$ tool servers, each potentially
executing in a different \tee on different hardware.
Composing attestation claims across this chain requires a form
of attestation transitivity that no current TEE architecture
natively supports.

Given that agent A has verified agent B's attestation and agent B
has verified agent C's, the conditions under which the user can
transitively trust C depend on the semantic content of the claims,
the delegation model, and whether any intermediary has been granted
re-attestation authority.
Blockchain-anchored approaches~\cite{blocka2a_2025} partially
address this through provenance ledgers, but a complete
attestation calculus for multi-hop agent chains, one that
jointly captures execution integrity, provenance, freshness,
revocation, delegation scope, and chain verification
efficiently, has not been proposed.
This is one of the more basic open problems in the space.

\subsection{TEE-Backed RAG and Vector Store Isolation}
\label{sec:open:memory}

Long-term vector stores are among the most sensitive assets
in an agentic system and are currently unprotected in all
deployed CC frameworks.
An attested, TEE-backed vector store, where retrieval
operations are gated by attestation, would close the memory
layer gap identified in Table~\ref{tab:threat_coverage}.
The design challenges include: efficient attestation of
per-query retrieval results without exposing the full index;
multi-user access control within a single TEE-resident store;
provenance and freshness of returned retrieval results; and
integration with existing vector database systems
such as Faiss~\cite{johnson_faiss_2021} and pgvector~\cite{pgvector_github}.

\subsection{CC-Aware Agent Communication Protocol Design}
\label{sec:open:protocols}

Current \mcp and A2A specifications contain no attestation
or confidentiality primitives at the protocol level.
AttestMCP~\cite{attestmcp_2026} proposes backward-compatible
extensions, but integration with TEE attestation at
the server end, enabling a client to verify that an
\mcp tool server is running in an attested enclave, remains
a protocol design problem without a standardized solution.
A CC-native agent communication protocol would treat attestation,
message authenticity, provenance, and freshness as first-class
primitives: server capability claims would be backed by hardware
attestation tokens, message authenticity would be bound to the
attested sender, provenance would survive multi-hop delegation,
and message confidentiality would extend to the \tee boundary.

\subsection{Side-Channel Leakage in Autoregressive Inference}
\label{sec:open:sidechannel}

LLM inference has a distinctive access pattern: autoregressive
token generation produces a predictable, sequential series of
memory accesses that may be observable to a co-tenant attacker
(\textbf{A2}) even within a TEE.
A February 2026 attack~\cite{tee_vuln_partial_2026} demonstrated
that even partial TEE shielding can be undermined through
cryptographic design flaws.
The broader SGX side-channel literature establishes the
attack vocabulary: controlled-channel attacks exploit page-fault
patterns observable by an untrusted OS~\cite{controlled_channel_sp_2015};
the Foreshadow attack recovers SGX enclave keys via transient
out-of-order execution~\cite{foreshadow_security_2018}; and
SgxPectre exploits speculative execution to leak enclave
secrets~\cite{sgxpectre_eurosp_2019}.
Side-channel analysis specific to transformer attention patterns,
KV-cache access, and speculative decoding has not been
systematically studied in TEE contexts, and translating this
prior attack vocabulary to LLM-specific memory-access patterns
is an open problem.

\subsection{GPU-TEE Performance at LLM Scale}
\label{sec:open:perf}

Reported GPU-TEE overhead of 4--8\% for \ngpu and 22\% for
CCA on-device inference is measured under specific workload
conditions~\cite{confidential_llm_perf_2025,cca_llm_eval_2025}.
Key unknowns include: overhead scaling with model size
(benchmarks have focused on the 1--70B range, with 405B+
models under-studied), latency behavior under concurrent
multi-agent workloads that share GPU resources, and
the CPU--GPU data-path overhead of TDX Connect under
sustained high-throughput agentic pipelines~\cite{intel_tdx_connect_2025}.
Comprehensive performance characterization across these
dimensions does not yet exist.

\subsection{Regulatory Alignment and Compliance}
\label{sec:open:regulatory}

The EU AI Act (Articles 9 and 17) and DORA (Articles 5--10)
impose explicit data-in-use protection requirements on
high-risk AI systems and financial infrastructure,
respectively~\cite{eu_ai_act_2024, dora_2022}.
A 2025 IDC study found that 77\% of organizations cite DORA
compliance as a driver for CC
adoption~\cite{idc_cc_2025}.
The regulatory question extends the technical one: how do attestation evidence, execution integrity,
provenance, and freshness become concrete deployment-assurance
artifacts?
In practice, a regulated agent deployment may need to show which model,
runtime, tool policy, and hardware configuration processed regulated
data; which parts of the workflow actually remained inside attested
trust boundaries; whether delegated tool calls preserved provenance and
policy compliance; and whether rollback-sensitive state or stale
attestation evidence could have invalidated those claims.

The current gap is that neither the CC literature nor regulatory
guidance specifies standard evidence bundles for this purpose.
Bridging it therefore requires both technical work
(machine-readable attestation evidence, deployment-assurance profiles,
and auditable provenance records) and policy work that maps those
artifacts onto the assurance expectations of the EU AI Act and DORA.

% ============================================================
\section{Conclusion}
\label{sec:conclusion}
% ============================================================

Agentic AI systems are being deployed at scale in regulated
industries, sensitive enterprise environments, and edge settings
where user data carries the highest privacy expectations.
Four principal takeaways emerge.

\noindent\textbf{First,} the surveyed literature suggests that confidential computing is
a useful, but partial, defense against the \emph{infrastructure}
threat layer: across inference, memory, tool mediation, and inter-agent
exchange, hardware isolation and attestation can prevent a privileged
operator from reading or tampering with sensitive state in ways that
software-only defenses cannot.

\noindent\textbf{Second,} a recurring pattern across the reviewed systems is selective
protection rather than wholesale enclosure.
TEESlice and CMIF protect sensitive inference paths, Omega places tool
policy and execution checks inside an attested runtime, AttestMCP
pushes message authenticity and provenance into protocol establishment,
and CAEC extends memory confidentiality to shared state.
That pattern points to a design lesson: the trust
boundary should be aligned with the layer where secrets, authority, and
cross-party provenance actually reside.

\noindent\textbf{Third,} the literature remains strongest for input confidentiality,
model confidentiality, and isolated execution integrity during
inference, but much weaker on persistent memory confidentiality,
compound attestation across delegation chains, standardized
support for message authenticity and provenance in \mcp and A2A, and
side-channel resilience under realistic LLM workloads.
Even the stronger evidence base is still largely inference-oriented, so
transfer to memory, action, and coordination remains only partial.

\noindent\textbf{Fourth,} the next advances are likely to be compositional rather than
purely architectural: attested memory services, protocol-native
capability and provenance mechanisms, GPU-aware performance
characterization, cross-boundary deployment patterns, and evidence
pipelines that turn attestation into usable deployment assurance.
If those pieces mature together, CC can become a more usable security
substrate on which safer agentic systems can be built.

\section*{Acknowledgements}
This research was supported by the UKRI Open Plus Fellowship
(EP/W005271/1), \textit{Securing the Next Billion Consumer Devices on the
Edge}.

% ============================================================
%  References
% ============================================================
\begingroup
\sloppy
\bibliographystyle{IEEEtran}
\bibliography{references}
\endgroup

\end{document}